\documentclass{aa} 
\usepackage{graphicx}
\usepackage{txfonts}
\usepackage{natbib}
\usepackage{ulem}
\usepackage{color}
\def\thco{$^{13}$CO}
\def\ceo{C$^{18}$O}
\def\hho{H$_2$O}
\def\hhoe{H$_2^{18}$O}
\def\hhos{H$_2^{17}$O}
\def\hh{H$_2$}

\def\ammo{NH$_3$}
\def\mic{$\mu$m}
\def\kms{km\,s$^{-1}$}

\def\pow#1#2{#1$\times$10$^{#2}$}

\def\ccm{cm$^{-3}$}
\def\msol{M$_{\odot}$}
\def\menv{$M_{\rm env}$}
\def\mvir{$M_{\rm vir}$}

\def\lbol{$L_{\rm bol}$}

\def\vinf{$V_{\rm inf}$}
\def\vlsr{$V_{\rm LSR}$}

\def\gtsim{{_>\atop{^\sim}}}
\def\ltsim{{_<\atop{^\sim}}}

\def\old#1{}
\def\older#1{}
\def\new#1{#1}
\def\newer#1{#1}
\begin{document}
\title{Multi-line Herschel/HIFI observations of water reveal infall motions and chemical segregation around high-mass protostars 
\thanks{\textit{Herschel} is an ESA space observatory with science instruments provided
by European-led Principal Investigator consortia and with important participation from NASA}}
\authorrunning{}
\titlerunning{Infall onto high-mass protostars}

\author{F.F.S. van der Tak \inst{\ref{sron},\ref{rug}} \and
           R.F. Shipman \inst{\ref{sron},\ref{rug}} \and
         T. Jacq \inst{\ref{bordeaux}} \and
       F. Herpin \inst{\ref{bordeaux}} \and
     J. Braine \inst{\ref{bordeaux}} \and
   F. Wyrowski \inst{\ref{mpifr}}
}

\institute{SRON Netherlands Institute for Space Research, Landleven 12, 9747 AD Groningen, The Netherlands; \email{vdtak@sron.nl} \label{sron} 
  \and Kapteyn Astronomical Institute, University of Groningen, The Netherlands \label{rug}
  \and Universit\'e de Bordeaux, France \label{bordeaux} 
  \and Max-Planck-Institut f\"ur Radioastronomie, Auf dem H\"ugel 69, 53121 Bonn, Germany \label{mpifr} }

\date{Submitted 6 July 2018 / Accepted 19 March 2019}

\abstract
{The physical conditions during high-mass star formation are poorly understood. 
Outflow and infall motions have been detected around massive protostellar objects, but their dependence on mass, luminosity, and age is unclear.
\new{In addition, physical conditions and molecular abundances are often estimated using simple assumptions such as spherical shape and chemical homogeneity, which may limit the accuracy.}}
{We aim to characterize the dust and gas distribution and kinematics of the envelopes of high-mass protostars. 
In particular, we search for infall motions, \new{abundance variations}, and deviations from spherical symmetry, \new{using Herschel data from the WISH program}.}
{We use HIFI maps of the 987~GHz \hho\ $2_{02}$--$1_{11}$ emission to measure the sizes and shapes of 19 high-mass protostellar envelopes.
To identify infall, we use HIFI spectra of the optically thin \ceo\ 9--8 and \hhoe\ $1_{11}$--$0_{00}$ lines.  
The high-$J$ \ceo\ traces the warm central material and redshifted \hhoe\ $1_{11}$--$0_{00}$ absorption indicates material falling onto the warm core. 
We probe small-scale \new{chemical differentiation} by comparing \hho\ 752 and 987 GHz spectra \new{with those of \hhoe}.}  
{Our measured radii of the \newer{central part of the} \hho\ $2_{02}$--$1_{11}$ emission are 30-40\% larger than the predictions from spherical \newer{envelope} models, and axis ratios are $<$2, which we consider good agreement. 
For 11 of the 19 sources, we find a significant redshift of the \hhoe\ $1_{11}$--$0_{00}$ line relative to \ceo\ 9--8.
The inferred infall velocities are 0.6--3.2 \kms, and estimated mass inflow rates range from \pow{7}{-5} to \pow{2}{-2} \msol/yr, with the highest mass inflow rates occurring toward the sources with the highest masses, and possibly the youngest ages. 
The other sources show either expanding motions or \hhoe\ lines in emission. 
The \hhoe\ $1_{11}$--$0_{00}$ line profiles are remarkably similar to the \textit{differences} between the \hho\ $2_{02}$--$1_{11}$ and $2_{11}$--$2_{02}$ profiles, suggesting that the \hhoe\ line and the \hho\ $2_{02}$--$1_{11}$ absorption originate just inside the radius where water evaporates from grains, typically 1000--5000 au from the center.
In some sources, the \hhoe\ line is detectable in the outflow, where no \ceo\ emission is seen.}  
{Together, the \hhoe\ absorption and \ceo\ emission profiles show that the water abundance around high-mass protostars has at least three levels: low in the cool outer envelope, high within the 100\,K radius, and very high in the outflowing gas.  
Thus, despite the small regions, the combination of lines presented here reveals systematic inflows and chemical information about the outflows.}


\maketitle

\section{Introduction}
\label{s:intro}

\begin{table*}[t]
\begin{flushleft}
\caption{Source sample.}
\label{t:sample}
\begin{tabular}{lllcccc}
  \hline \hline
\noalign{\smallskip}
Source$^a$  & RA  (J2000.0) & Dec   & $L_{\rm bol}$ & $d$ &  $M_{\rm env}$ & Distance \\
&    hh mm ss.s         & $^0$ $'$ $''$   & $L_\odot$      & kpc  &  $M_\odot$      & Reference \\
\noalign{\smallskip}
\hline
\noalign{\smallskip}
{\bf mid-IR-quiet HMPOs}$^b$&    &    &        &    &       \\
\noalign{\smallskip}
IRAS 05358+3543 & 05 39 13.1    &   +35 45 50     & \pow{6.3}{3}     & 1.8   &  142 & (1) \\
IRAS 16272--4837 & 16 30 58.7    & $-$48 43 55    & \pow{2.4}{4}    & 3.4  &  2170 & (1) \\
NGC 6334I(N)        & 17 20 55.2    & $-$35 45 04    & \pow{1.1}{3}  & 1.3  &  2237 & (5) \\
W43 MM1                & 18 47 47.0    & $-$01 54 28   & \pow{1.8}{4}    & 5.5  &  5992 & (2) \\
DR21(OH)               & 20 39 00.8    &   +42 22 48   & \pow{1.3}{4}     & 1.5  &  472 & (1) \\
\noalign{\smallskip}
{\bf mid-IR-bright HMPOs}&    &    &        &    &       \\
\noalign{\smallskip}
W3 IRS5                   & 02 25 40.6    &   +62 05 51   & \pow{1.7}{5}   & 2.0 & 424  & (1) \\
IRAS 18089--1732 & 18 11 51.5    & $-$17 31 29     & \pow{1.3}{4}     & 2.3  & 172 & (1) \\
W33A                        & 18 14 39.1    & $-$17 52 07   & \pow{4.4}{4}   & 2.4  & 700 & (1) \\
IRAS 18151--1208 & 18 17 58.0    & $-$12 07 27      & \pow{2.0}{4}     & 2.9 &  153 & (1) \\
AFGL 2591              & 20 29 24.7    &   +40 11 19      & \pow{2.2}{5}     & 3.3  & 363 & (1) \\
\noalign{\smallskip}
{\bf Hot Molecular Cores}&    &    &        &    &      \\
\noalign{\smallskip}
G327$-$0.6         & 15 53 08.8    & $-$54 37 01      & \pow{4.4}{4}     & 3.1  & 1804 & (6) \\
NGC 6334I          & 17 20 53.3    & $-$35 47 00      & \pow{1.5}{5}      & 1.3 & 439 & (5) \\
G29.96$-$0.02   & 18 46 03.8    & $-$02 39 22      &  \pow{2.7}{5}       & 5.3  & 599 & (2) \\
G31.41+0.31       & 18 47 34.3    & $-$01 12 46      & \pow{8.8}{4}      & 4.9  & 1142 & (2) \\
\noalign{\smallskip}
{\bf Ultracompact H{\sc II} Regions}&    &    &        &    &       \\
\noalign{\smallskip}
G5.89$-$0.39 (W28A) & 18 00 30.4    & $-$24 04 02      & \pow{5.1}{4}      & 1.3  & 140 & (1)  \\
G10.47+0.03                 & 18 08 38.2    & $-$19 51 50    & \pow{8.1}{5}    & 8.6  & 2568 & (3) \\
G34.26+0.15                 & 18 53 18.6    &   +01 14 58     & \pow{7.5}{4}     & 1.6  & 421 & (4) \\
W51N-e1                       & 19 23 43.8    &   +14 30 26     & \pow{1.1}{5}     & 5.4 & 5079 & (7) \\
NGC 7538-IRS1           & 23 13 45.3    &   +61 28 10     & \pow{1.3}{5}     & 2.7  &  433 & (1) \\
\noalign{\smallskip}
\hline
\end{tabular}
\tablefoot{$^a$: The text uses "short" source names, which is the part preceding the $+$ or $-$ sign. $^b$: \newer{High-Mass Protostellar Objects}. \\
References: (1) See \citet{vandertak2013}; (2) \citet{zhang2014}; (3) \citet{sanna2014}; (4) \citet{kurayama2011,xu2016}; (5) \citet{wu2014}; (6) \citet{wienen2015}; (7) \citet{sato2010}.}
\end{flushleft}  
\end{table*}

High-mass stars ($>$8\,\msol) play a key role in the evolution of their host galaxies, but their formation is poorly understood, especially for masses $>$20\,\msol.
The leading models of high-mass star formation involve infall from a dense protostellar core, and accretion onto the protostar via a circumstellar disk \citep{tan2014,motte2017}.
While rotating disks have been detected around young B-type \citep{sanchez-monge2013,beltran2016} and O-type \citep{johnston2015,cesaroni2017} protostars, the exact manner in (and rate at) which material is gathered from the surroundings is still a matter of debate.

In the `monolithic collapse' model, a massive dense core collapses under its own gravity and forms a (cluster of) protostar(s), much like the low-mass case.
This picture is supported by observations of massive collimated outflows from high-mass protostars \citep{beuther2002}. 
In the alternative `competitive accretion' model, the accreting protostellar core is replenished from the surroundings.
Evidence supporting this model comes for example from observations of extended contracting motions in pre-protocluster regions \citep{pillai2011}.
Possibly both models are valid under different conditions, or combination models need to be developed \citep{peters2011}.
To constrain such models, observations of suitable tracers are essential.

Large-scale ($\sim$0.1\,pc) infall motions have been detected toward high-mass star-forming regions in ground-based submillimeter-wave molecular emission line maps \citep{motte2003,peretto2006}, in redshifted \ammo\ line absorption at centimeter wavelengths \citep{sollins2005,beltran2006}, and recently in SOFIA \ammo\ spectra \citep{wyrowski2012,wyrowski2016}.
Searches for infall in unbiased \newer{selections} from catalogues of high-mass star-forming regions confirm the ubiquity of such motions \citep{fuller2005,klaassen2007,he2015,cunningham2018}.

The water molecule appears to be a promising tracer of infall motions in low-mass star-forming regions \citep{mottram2013}, \new{and \citet{sanjose2016} linked water observations between low- and high-mass star-forming regions.}
Spectra of low-$J$ line emission toward high-mass objects often exhibit inverse P~Cygni profiles \citep{vandertak2013}, which have been modeled successfully as infall, using spherical Monte Carlo models \citep{herpin2016}.
Stronger evidence comes from maps of the luminous mini-starburst region W43 in low-energy \hho\ and \hhoe\ lines \citep{jacq2016}: extended \hhoe\ absorption which is redshifted with respect to the \thco\ 10-9 emission clearly indicates infall motions.

The \hhoe\ $1_{11}$--$0_{00}$ ground-state absorption towards W43-MM1 is remarkable because its shape closely matches that of the central absorption feature in the \hho\ $2_{02}$--$1_{11}$ excited-state emission line \citep{jacq2016}. 
This resemblance strongly suggests that the two lines originate in the same gas, which is curious given their different excitation energies (101 vs 0\,K). 
In order to understand the similarity of these line profiles, this paper explores whether the same effect is seen in other high-mass protostars.

Another puzzle in previous observations of \hho\ lines towards high-mass protostars concerns their line shape \citep{vandertak2013}. 
The \newer{profiles} show narrow line cores from the protostellar envelopes, and broad line wings from the outflows, but the wings are much more pronounced at red- than at blueshifted velocities, and often the blueshifted wings are nearly or entirely missing from the profiles.
This asymmetry cannot be due to continuum absorption (e.g. by a disk) which would preferentially affect background gas (i.e., receding velocities). 
Special geometrical configurations may explain individual cases, but not a sample of many sources.
One possibility is that the \hho\ excitation temperature is close to the brightness temperature of the background, so that no net \new{line emission or absorption appears} in the spectra.
To explore the origin of the asymmetry, this paper explores its dependence on line properties such as excitation energy and critical density.

This paper uses multi-line maps and spectra of \hho\ and \hhoe\ lines toward a sample of high-mass protostars to explore the gas distribution and dynamics.
In particular, we compare \hhoe\ line profiles to those of \ceo\ in order to search for velocity shifts due to infall motions.
Furthermore, we use \hho\ maps to measure the sizes of the protostellar envelopes, and to test the assumption of spherical symmetry in previous analyses of pointed spectra.
Section~\ref{s:obs} describes our observations, and
Section~\ref{s:res} presents the resulting maps and spectra.
Section~\ref{s:disc} compares our derived infall rates with previous observations and with models, and searches for trends with basic source parameters. 
Finally, Section~\ref{s:concl} describes our conclusions. 

\begin{table*}
\caption{\new{Observation log for the \hhoe\ pointed observations and \hho\ 987~GHz maps}.}            
\label{t:obslog}      
\centering                    
\begin{tabular}{llccc}   
\hline\hline             
Source & Species  &  ObsID$^a$ & $t_{int}$$^b$ & rms$^c$ \\
              &                   &             &    [s]               &[mK] \\
\hline
 IRAS05358 &  \hhoe\    &  206124, 206126 &  3566 & 20   \\
   & \hho\      & 204508 &  5.86 & 481 \\
\hline
 IRAS16272 &  \hhoe\    &  214417, 214419 & 3455 & 20   \\
  & \hho\      & 203166 &  5.86 & 473 \\
\hline
 NGC6334I(N) &  \hhoe\    & 206383 &  2965  & 22   \\
  & \hho\      & 204523 &  5.86 & 481 \\
 \hline
W43--MM1 &  \hhoe\    &  191670, 207372 &  3566 & 20   \\
  & \hho\      & 215899 &  5.86 & 524 \\
\hline
 DR21(OH) &  \hhoe\    &  194794, 197974 & 3566 & 20   \\
  & \hho\      & 210042 &  7.86 & 396  \\
 \hline
W3IRS5 &  \hhoe\    &  191658, 201591 &  3566 & 20   \\
  & \hho\      & 203160 &  5.86 & 498 \\
  \hline
 IRAS18089 &  \hhoe\    & 229882, 229883 & 3455 & 20   \\
 & \hho\      & 218210 &  5.86 & 557 \\
\hline
W33A &  \hhoe\    &  191638, 208086  &  3566 & 20   \\
 & \hho\      & 215902 &  5.86 & 465 \\
\hline
 IRAS18151 &  \hhoe\    &  229880, 229881 & 3455   & 20   \\
  & \hho\      & 218212 &  5.86 & 659  \\
\hline
AFGL2591 &  \hhoe\    &  194795, 197973  & 3566 & 20   \\
  & \hho\      & 210038 &  5.86 & 446 \\
\hline
G327 &  \hhoe\    &  214422, 214423, 214425, 214426 &  3428   & 21   \\
  & \hho\      & 203169 &  5.86 & 490 \\
 \hline
NGC6334I &  \hhoe\    &  206385 &  2965  & 22 \\
 & \hho\      & 204522 &  5.86 & 486 \\
 \hline
G29.96 &  \hhoe\    &  191668, 191669, 229875, 229876&  3700   & 20   \\
  & \hho\      & 207655 &  5.86 & 475  \\
\hline
G31.41 &  \hhoe\    &  191671, 191672, 229873, 229874 &  3700   & 20   \\
 & \hho\      & 207654 &  5.86 & 477 \\
\hline
G5.89 &  \hhoe\    &  229888, 229889, 229890, 229891 &  3148   & 21   \\
 & \hho\      & 218201 &  5.86 & 575 \\
\hline
G10.47 &  \hhoe\    &  229884, 229885, 229886, 229887 &  3148  & 21   \\
 & \hho\      & 218208 &  5.86 & 476 \\
\hline
G34.26 &  \hhoe\    &  191673, 191674, 229871, 229872 &  3700   & 20   \\
 & \hho\      & 207652 &  5.86 & 493 \\
\hline
W51 &  \hhoe\    &  194801, 194802, 207384, 207385 &  3420   & 20   \\
 & \hho\      & 207651 &  5.86 & 485  \\
\hline
NGC7538IRS1 &  \hhoe\    & 191663, 191664, 197976 &  3569   & 20   \\
 & \hho\      & 203161 &  5.86 & 479 \\
\hline
\end{tabular}
\tablefoot{$^a$ The leading 1342 has been omitted. 
$^b$ For pointed observations, the integration time is for the total spectra, i.e. all ObsIDs added. For maps, the integration time is per observed position.
$^c$ The rms is the noise in $\delta \nu=1.1$MHz. For pointed observations, the integration time is for the total spectra, i.e. all ObsIDs added.}
\end{table*}

\section{Observations}
\label{s:obs}

\subsection{Source sample}
\label{ss:sample}

As part of the guaranteed time program WISH (Water In Star-forming regions with Herschel; \citealt{vandishoeck2011}), we have selected 19 regions of high-mass star formation for observation in lines of \hho\ and its isotopes with the HIFI instrument \citep{degraauw2010} on ESA's Herschel Space Observatory \citep{pilbratt2010}.
The sources were selected to cover wide ranges in bolometric luminosity, mid-infrared brightness, and circumstellar mass, and to include regions with hot molecular cores and ultracompact HII regions; see \citet{vandertak2013} for details.
Table~\ref{t:sample} presents the source sample, where distances have been updated following \citet{koenig2017}, and luminosities and masses have been scaled assuming a simple $d^2$ dependence.

Most of the updated distances are direct determinations using trigonometric maser parallax observations.
The near kinematic distance for G327 seems to be broadly accepted in the recent literature.
Only the case of G31.41 is more complicated.
The commonly used distance for G31.41 is 7.9 kpc, based on its radial velocity from the Sun and position on the sky, coupled with a Galactic rotation model \citep{churchwell1990}.
However, such kinematically derived distances can be off by factors of $\gtsim$2 in either direction; AFGL 2591 and W33A are cases in point \citep{rygl2012,immer2013}.
Alternatively, G31.41 may be associated with the W43-Main cloud complex, as suggested by position-velocity diagrams of the molecular gas in the surroundings \citep{nguyen-luong2011}. 
For W43-Main, two distance estimates exist which are based on VLBI observations of maser parallax \new{(see also \citealt{beltran2018})}. 
\citet{reid2014} report a distance of 4.9~kpc to the W43-Main core, while \citet{zhang2014} report distances to 5 maser spots with distances ranging from 6.21 to 4.27 kpc.
Given this large spread, we adopt a distance of \new{4.9~kpc} for G31.41 in this paper, and recommend a specific maser parallax study of G31.41 itself.

\subsection{Data acquisition and reduction}
\label{ss:data}

Maps of the \hho\ $2_{02}$--$1_{11}$ line at 987.927~GHz (hereafter 987 GHz) were taken with HIFI Band 4a. 
The maps are 1$'$ on the side, and were taken in on-the-fly (OTF) observing mode.
The backend was the acousto-optical Wide-Band Spectrometer (WBS) which provides a bandwidth of 4$\times$1140 MHz (1200 \kms) at a resolution of 1.1 MHz (0.3 \kms).
Table~\ref{t:obslog} presents a detailed observation log including integration times; system temperatures were around 340\,K. 
The FWHM beam size at this frequency is 22$''$ \citep{roelfsema2012}, which corresponds to 0.14--0.92\,pc at the distances of our sources. 
The maps thus cover at least part of the protostellar outflows, while the beam resolves the protostellar envelopes, but not any possible disks.  

Spectra of the \hhoe\ $1_{11}$--$0_{00}$ line at 1101.698 GHz (hereafter 1101 GHz), the \hho\ $2_{11}$--$2_{02}$ line at 752.033 GHz  (hereafter 752 GHz),  the \thco\ 10--9 line at 1101.34976 GHz, and the \ceo\ 9--8 line at 987.560 GHz were obtained toward the same sources with HIFI, using the Double Beam Switch observing mode with a chopper throw of 3$'$.
The \ceo\ and \thco\ lines have been observed in the same tuning as the \hho\ $2_{11}$--$2_{02}$ and the \hhoe\ $1_{11}$--$0_{00}$ lines, respectively, and thus share the same ObsIDs.
Table~\ref{t:obslog} lists the integration times of the spectra; system temperatures were around 200 and 390\,K for the 752 GHz and $\sim$1\,THz lines, respectively. 
The pointed 987 and 752 GHz spectra have been presented before by \citet{sanjose2016}; \newer{the \thco\ and \ceo\ spectra were presented in \citet{sanjose2013}.}
The DBS spectra have higher noise per second of integration than the maps at the same frequency, which represents the noise `penalty' to be paid for stabilizing the system by differencing two reference positions in the DBS observing mode.
For the \hhoe\ and \ceo\ lines, the beam size of 20--22$''$ is very similar to that of the 987~GHz maps, which permits a direct comparison of the results.
The beam size of the 752 GHz observations is 28$''$.

The data are Herschel/HIFI standard products \citep{shipman2017} with further processing performed in the \textit{Herschel} Interactive Processing Environment (HIPE; \citealt{ott2010}) version 15; 
further analysis was carried out in the CLASS\footnote{\tt http://www.iram.fr/IRAMFR/GILDAS} package, version of December 2015 or later. 
Raw antenna temperatures were converted to $T_{\rm mb}$ scale using a main beam efficiency of 63\% for both frequencies around 1\,THz and 64\% for the 752\,GHz line\footnote{\tt https://www.cosmos.esa.int/web/herschel/ legacy-documentation-hifi-level-2}, and linear baselines were subtracted. 
After inspection, the data from the two polarization channels were averaged to obtain the rms noise levels reported in Table~\ref{t:obslog}. 
The absolute calibration uncertainty of HIFI Bands~3 and~4 is estimated to be 10-15\%,
but the relative calibration between lines in the same spectrum should be much better, which is relevant for \ceo\ and \thco.

\section{Results}
\label{s:res}

\subsection{Line profiles of \hhoe, \thco, and \ceo\ }
\label{ss:profiles}

\begin{figure*}[tb]
\centering
\includegraphics[width=9cm,angle=0]{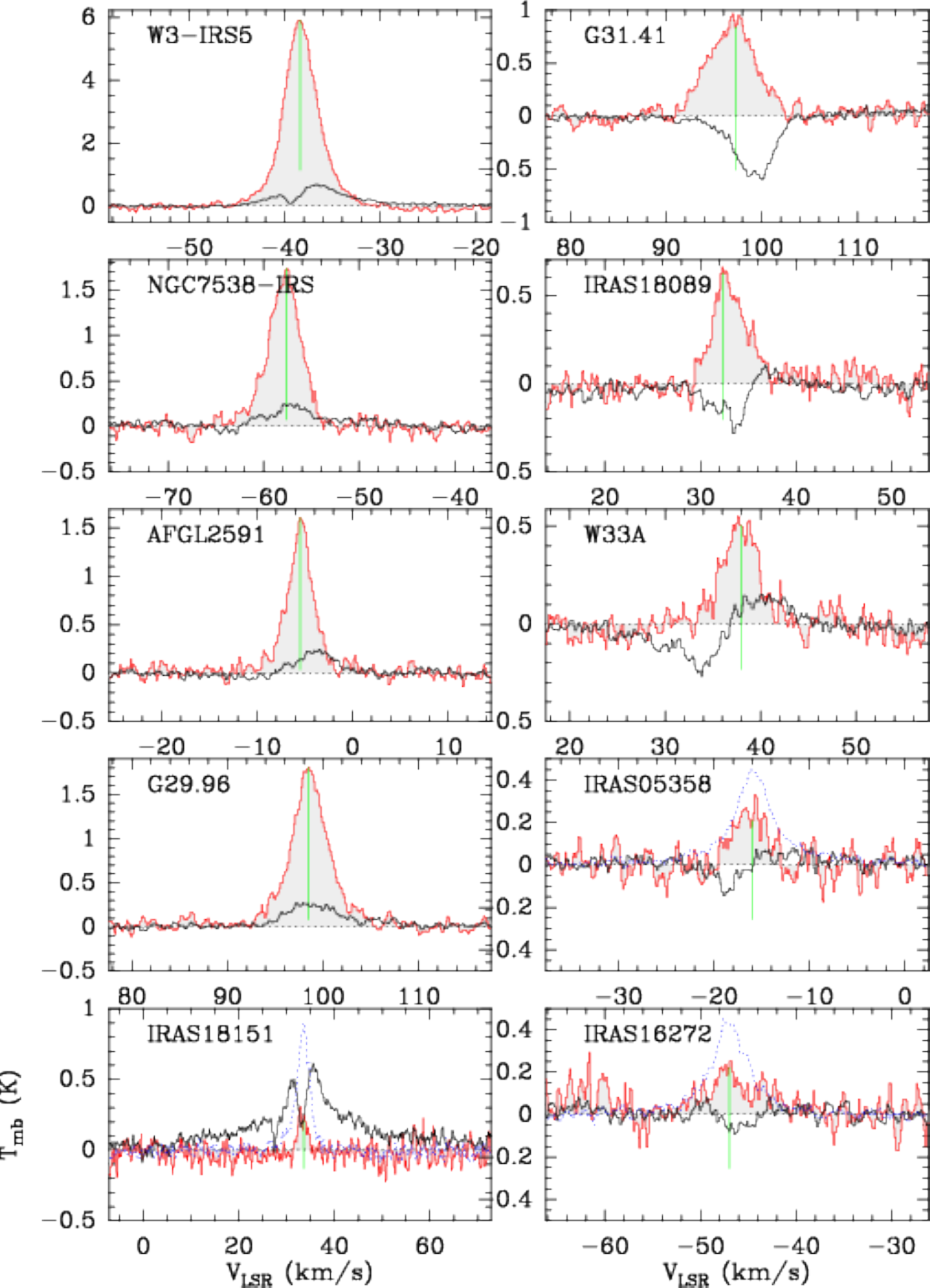}
\includegraphics[width=9cm,angle=0]{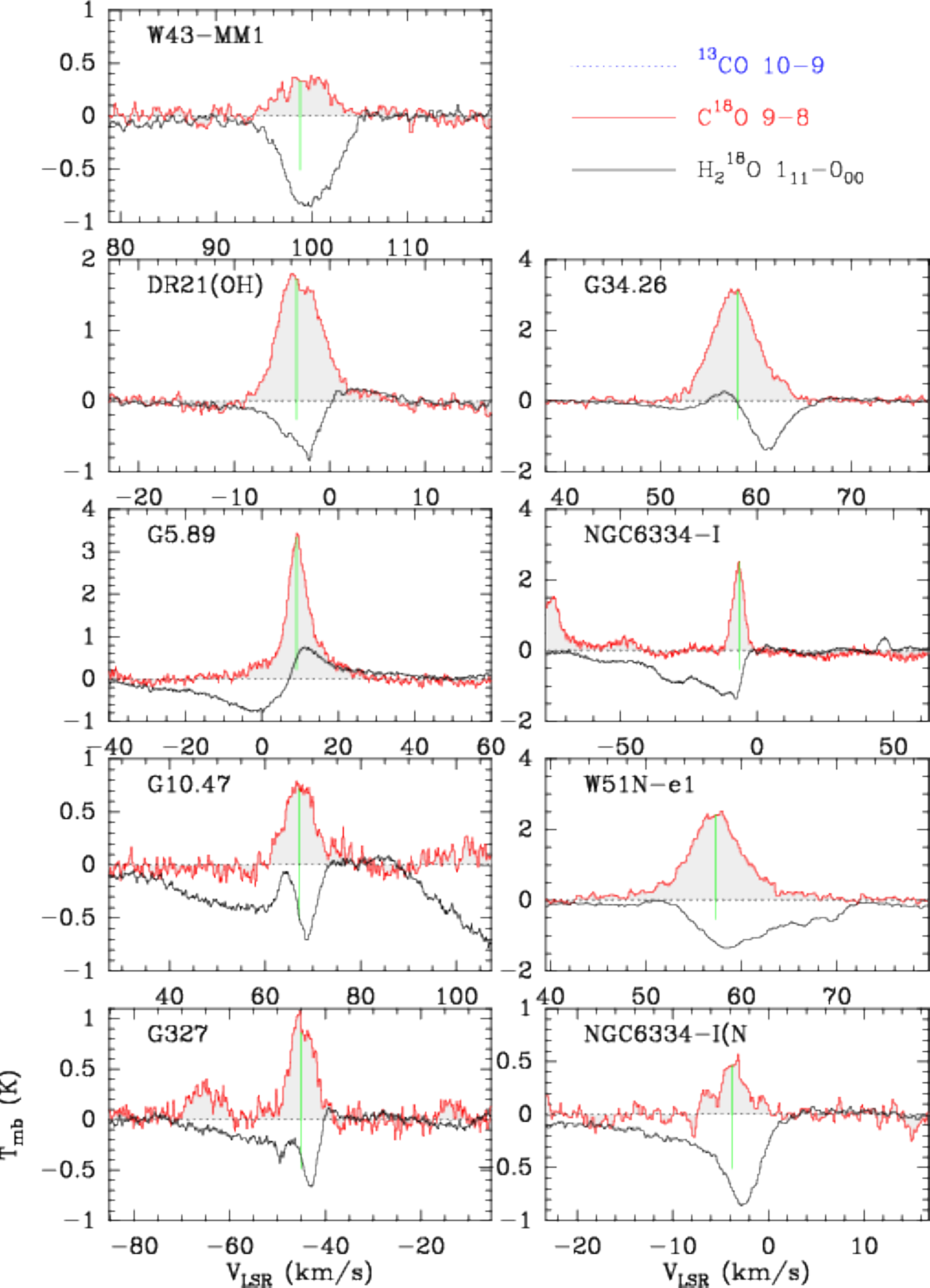}
\caption{Line profiles of \hhoe\ $1_{11}$--$0_{00}$ (black) and \ceo\ 9--8 (red) toward our 19 sources. 
The vertical green line denotes the \ceo\ velocity in Table~\ref{t:pars}. 
For IRAS 18151, we show \hho\ $1_{11}$--$0_{00}$ \new{instead of the \hhoe\ line} which is not detected.
For IRAS 18151, IRAS 16272 and IRAS 05358, the blue dotted spectrum is \thco\ 10--9 as \ceo\ is weak or noisy.
The dip in the G10.47 spectrum at \vlsr$>$80\,\kms\ is an artifact from the image sideband.
}
\label{f:profs}
\end{figure*} 

Figure~\ref{f:profs} shows the observed velocity profiles of the \hhoe\ $1_{11}$--$0_{00}$ and \ceo\ 9--8 lines. 
For IRAS 18151, we show the \hho\ $1_{11}$--$0_{00}$ line as the \hhoe\ line is not detected. 
For IRAS 05358, IRAS 16272, and IRAS 18151, the \ceo\ 9--8 line is weak, so we use the \thco\ 10--9 line to measure velocities.
\new{For the other sources, the data indicate substantial optical depth in the \thco\ 10--9 line, so we prefer \ceo\ 9--8 as velocity standard.}

While the \ceo\ (or \thco) lines appear purely in emission for all sources, the \hhoe\ (or \hho) profiles show absorption, in some cases mixed with emission.
Despite this difference, the peak of the \hhoe\ absorption is seen to lie close to the peak of the \ceo\ (or \thco) emission, but at a measureable velocity offset.
In most cases, the \hhoe\ absorption peak is significantly redshifted from the \ceo\ (or \thco) emission peak, by 0.6--3.2\,\kms.
Table~\ref{t:pars} reports the peak velocities of the \hhoe\ absorption and \ceo\ (or \thco) emission, as estimated directly from the HIFI spectra.
\new{We estimate the uncertainty on these velocities to be $\approx$0.3\,\kms. In some cases, no blueshifted absorption or no absorption at all is seen.}

The \ceo\ 9--8 line has a relatively high upper level energy (237\,K) and critical density (\pow{7.7}{5}\,\ccm), using spectroscopy from \citet{endres2016} and collision data from \citet{yang2010}, as provided on the LAMDA database \citep{schoeier2005}.
Since in addition the \ceo\ abundance is likely to be low ($\sim$10$^{-6}$), this line should be an optically thin tracer of the warm dense gas close to the central protostar.
For the 3 sources with weak \ceo\ 9--8 emission, this argument seems to hold also for the \thco\ 10--9 line, presumably due to a low envelope mass.
These 3 sources are not the lowest-luminosity cases in our study, so the low envelope mass \newer{and weak \ceo\ emission} may be an evolutionary effect.
The appearance of the \hhoe\ absorption at redshifted velocities thus implies infalling motions in the gas surrounding the dense warm cores seen in \ceo\ 9--8 and/or \thco\ 10--9 emission.
The velocity difference between the \ceo\ and \hhoe\ lines indicates approximate infall speeds between 0.6\,and 3.2\,\kms, although these values represent line-of-sight averages.

For the sources W3~IRS5, W33A, NGC 6334I, and IRAS 05358, the \hhoe\ absorption peak is blueshifted from the \ceo\ emission peak, suggesting expanding motions. 
The line profiles toward G5.89 and G10.47 are complex, with a mixture of infall and expansion. These two sources are not included in the analysis below.

We emphasize the importance of using a precise velocity standard, in this case the \ceo\ 9--8 line, for the detection of infall motions. 
The \ceo\ velocities in Table~\ref{t:pars} differ from the ground-based values (\citealt{vandertak2013}, Table~1; \new{\citealt{vandishoeck2011}}) by up to 1~\kms, which shows that velocity precision is often limited by source inhomogeneities, rather than by spectral resolution or other instrumental parameters.

Wyrowski et al. (\citeyear{wyrowski2012},\citeyear{wyrowski2016}) have used SOFIA to measure the \ammo\ $3_2^+$-$2_2^-$ line at 1810.379\,GHz toward several of our sources. 
They report redshifted absorption toward W43\,MM1, G327, G31.41, and G34.26, implying infall, and blueshifted absorption toward W33A and G5.89, implying expansion. 
These results agree qualitatively with ours, and their measured velocities are similar to those reported here.

Toward G34.26, \citet{hajigholi2016} have measured infall through multi-line \ammo\ line observations with HIFI, and found two infall components with velocities of 2.7 and 5.3 \kms. 
The ground-state \hhoe\ and \ammo\ lines presented here and by Wyrowski et al.\ only probe the lower-velocity of these components, which may mean that the higher-velocity component mostly arises in very warm and dense gas in close proximity to the protostar.
This result is consistent with a scenario where the infall velocity of the gas increases as it approaches the protostar.

\begin{table*}[t]
\caption{Measured velocities and derived \new{infall} rates.}
\label{t:pars}
\begin{tabular}{llllcccc}
  \hline \hline
\noalign{\smallskip}
Source             & $V$(\ceo) &  $V$(\hhoe) & \vinf$^a$ & $\dot{M}_{acc}$ $^b$\\
                        & \kms\ & \kms\ & \kms\ & 10$^{-3}$ \msol/yr \\
\noalign{\smallskip}
\hline
\noalign{\smallskip}
IRAS 05358         &--16.0$^d$  & --18.5 & +2.5 & -0- \\
IRAS 16272         & --47.0 & --46.3 & --0.7 & 1.28 \\ 
NGC 6334 I(N)    & --3.8 & --2.7 & --1.1 & 4.75 \\ 
W43 MM1            & +98.8 & +99.4 & --0.6 & 6.06 \\  
DR21(OH)           & --3.5 & --2.2 & --1.3 & 1.86  \\ 
W3 IRS5              & --38.4 & --39.9 & +1.5 & -0- \\ 
IRAS 18089         & +32.4  & +33.8 & --1.4 & 3.28 \\ 
W33A                   & +38.0 & +33.8 & +4.2 & -0- \\ 
IRAS 18151        & +33.6$^d$ & +33.4$^e$  & --0.2 & 0.07 \\ 
AFGL 2591$^c$  & --5.5 & -0- & -0- & -0- \\ 
G327-0.6             & --45.0 & --43.1 & --1.9 & 4.19 \\ 
NGC 6334I          & --6.5 & --7.9 & +1.4 & -0- \\ 
G29.96$^c$         & +98.5 & -0- & -0- & -0- \\ 
G31.41                 & +97.3 & +99.3 & --2.0 & 9.85 \\  
G5.89$^c$           & +9.3 & -0- & -0- & -0- \\  
G10.47                 &  +67.3 & +68.7 & --1.4 & 7.15 \\ 
G34.26                 & +58.0 & +61.2 & --3.2 & 18.9 \\ 
W51N-e1             & +57.3 & +58.9 & --1.6 & 14.2 \\ 
NGC 7538$^c$    & --57.6 & -0- & -0- & -0- \\
\noalign{\smallskip}
\hline
\end{tabular}
\tablefoot{$^a$: \vinf\ =  $V$(\ceo) $-$ $V$(\hhoe).
$^b$: Zero denotes lack of (blueshifted) absorption w.r.t. the source.
$^c$: \hhoe\ in emission without clear absorption.
$^d$: From \thco.
$^e$: From \hho. 
}
\end{table*}


\subsection{Maps of \hho}
\label{ss:maps}

\begin{figure*}
\includegraphics[width=18cm]{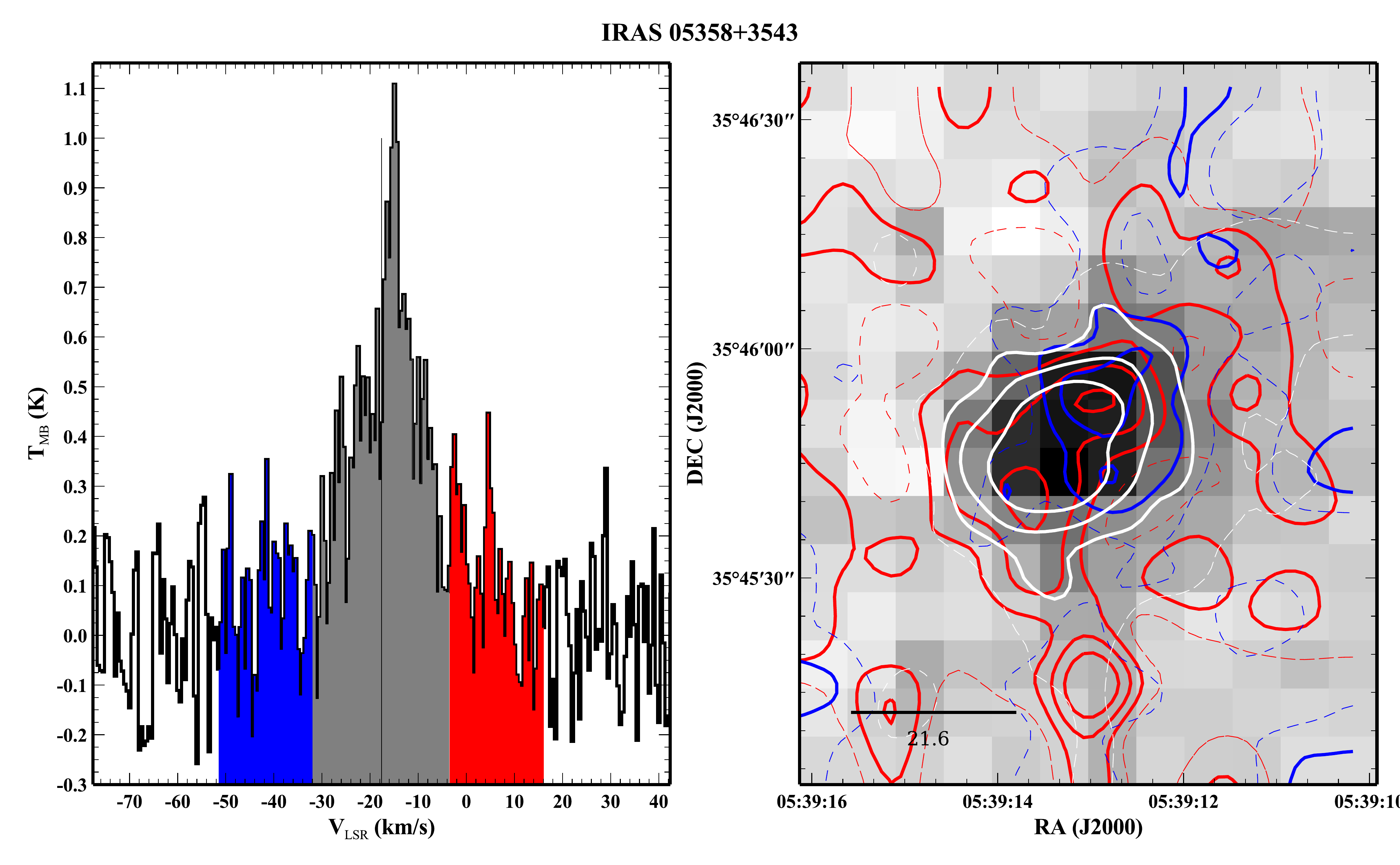}
\caption{Map of the velocity-integrated emission in the \hho\ 987~GHz line for IRAS 05358. 
White contours and grayscale image denote velocity-integrated emission over the range denoted by the grey area in the spectrum in the left panel. 
Red and blue contours denoted red- and blue-shifted emission, denoted by the red and blue areas in the left panel. 
The red and blue maps were created by integrating the 987 GHz data cube over a range of 1 FWHM below and above the \vlsr\ of the envelope, denoted by the vertical black line in the spectrum.  The integration ranges are offset by 1/2 FWHM from the \vlsr\ to avoid confusion with absorption features.
The lowest contour (at the 1$\sigma$ level) is drawn dashed, the others (in multiples of $\sigma$) are drawn solid. 
The bar in the bottom left corner denotes the HIFI beam size.}
\label{f:firstmap}
\end{figure*} 

\begin{figure*}
\includegraphics[width=18cm]{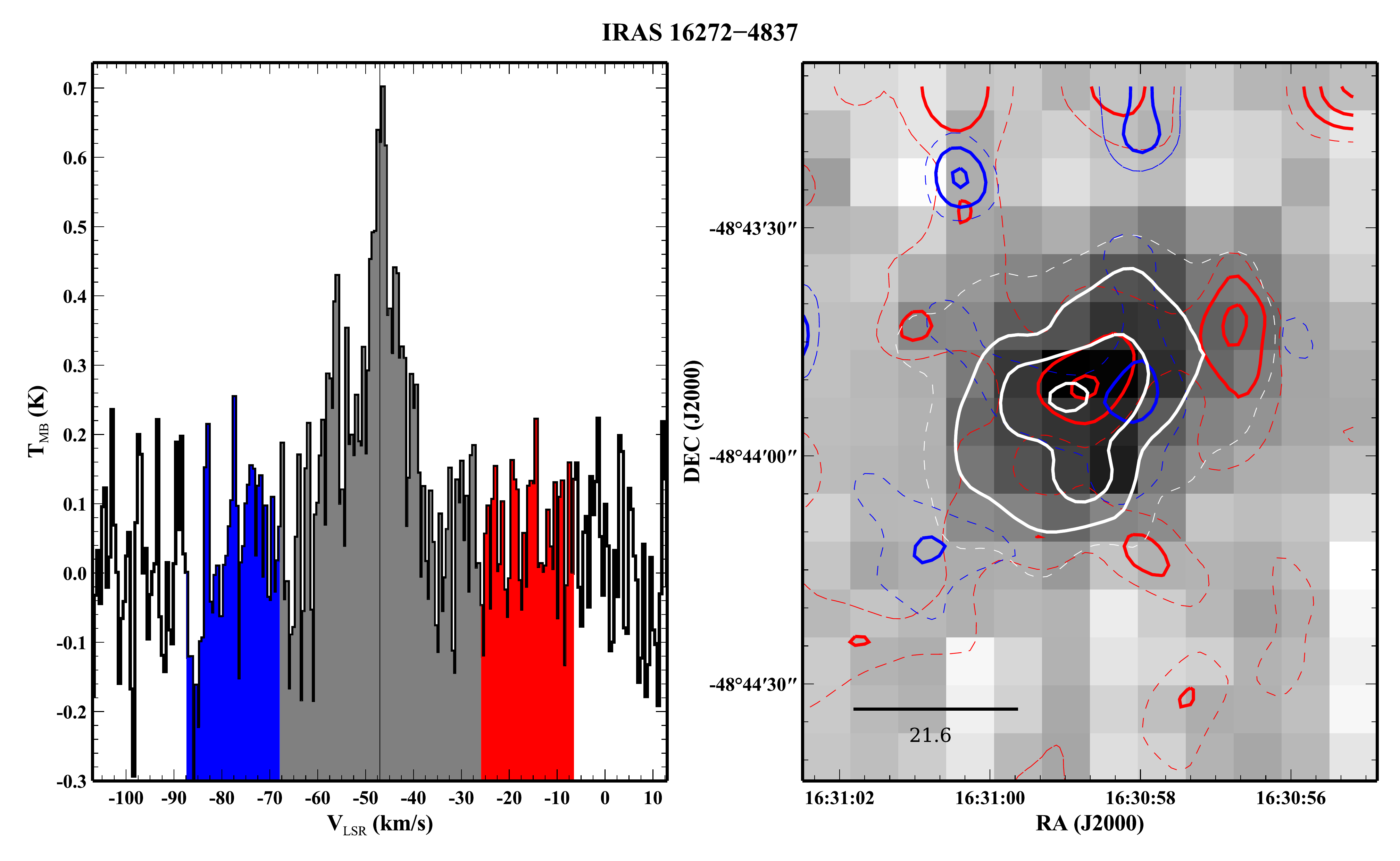}
\caption{As Figure~\ref{f:firstmap}, for IRAS 16272.}
\label{f:secondmap}
\end{figure*}

\begin{figure*}
\includegraphics[width=18cm]{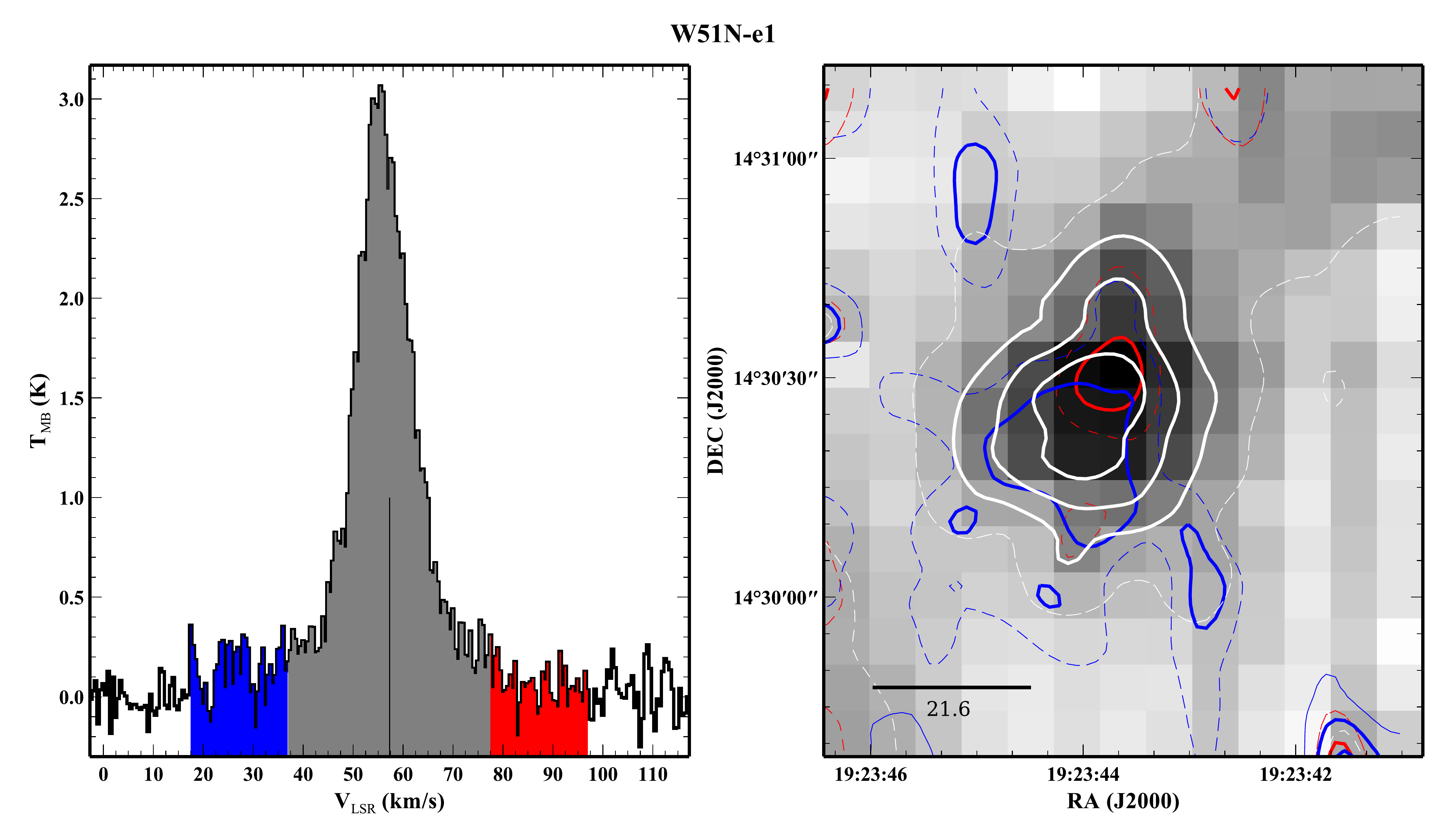}
\caption{As previous figure, for W51N.}
\label{f:lastmap}
\end{figure*}

Figures~\ref{f:firstmap} and \ref{f:secondmap}--\ref{f:lastmap} show our maps of the \hho\ 987~GHz line emission. 
The \newer{greyscale and white contours denote the line core}, while the blue and red contours correspond to the blue- and redshifted line wings (see the caption for details).
The emission is seen to be compact (except G31.41), mildly elongated, and not to depend much on velocity interval (except NGC 6334I).
For IRAS 18151, the emission is too weak to assess its morphology.
The map of NGC 7538 \new{is not shown as it} suffers from mispointing, so that only limits on the emission size and shape can be obtained.

The observed morphology of the 987~GHz emission does not appear to depend much on velocity interval (Figs.~\ref{f:firstmap} \& \ref{f:secondmap}--\ref{f:lastmap}).
\new{This contrasts with the} low-$J$ CO emission from our sources, which shows a clear bipolar morphology, especially at velocities away from line center (see references in \citealt{vandishoeck2011}).
We conclude that the bulk of the warm dense gas in the outflow as traced by the \hho\ 987~GHz line is confined to a small volume ($\ltsim$20$''$) from the source, unlike the outflow gas at lower temperature and density traced by low-$J$ CO lines.

We have measured the size of the 987~GHz emission by fitting a two-dimensional Gaussian plus a background offset to the images in Figs.~\ref{f:firstmap}  \& \ref{f:secondmap}--\ref{f:lastmap}.
Table~\ref{t:sizes} reports the resulting radii, which have been deconvolved assuming that the source and beam profiles add in quadrature.
The measured sizes \newer{of the \hho\ emission} are $\sim$2$\times$ smaller than the values measured in high-$J$ CO lines with PACS (\citealt{karska2014}; Kwon et al, submitted), and 2--3 times smaller than the sizes of the submm\older{/THz} dust emission measured with \newer{JCMT/SCUBA and APEX/LABOCA} \citep{vandertak2013}.
Evidently, the \hho\ emission traces warm dense gas close to the protostars.

Comparing the major and minor axis values in Table~\ref{t:sizes}, we see that the \hho\ emission is close to spherical in most cases, with axis ratios between 1.1 and 1.4.
We conclude that protostellar envelopes dominate the emission, without any evidence for flattening or elongation due to rotation or bipolar outflows.

Table~\ref{t:sizes} compares the observed shape of the \hho\ 987~GHz emission to the predictions from radiative transfer models, \newer{assuming a constant \hho\ abundance, following \citet{herpin2016}.}
These predictions are fits to multi-line \hho\ (and isotopic) spectra from HIFI, using the physical structure models from \citet{vandertak2013}.
\newer{The predicted size is seen to be 30--40\% larger than the observed size for most sources, which we consider good agreement given the simplifying assumption of spherical symmetry in the models.}
Only for the sources W3~IRS5 and W43~MM1, the predicted size is 2--4 times smaller than the observed size.
As with the axis ratios, this may be due to outflows contributing to the emission.
Furthermore, the models assume a single central source, whereas interferometric images of our objects often show multiple cores at the center \citep[e.g.,][]{hunter2014,brogan2016,izquierdo2018}.

The line intensities in the maps are typically 70-80\% of the values reported from pointed observations at the same position.
This difference is as expected from the 4\% larger beam size due to the OTF observing mode and the spatial regridding, assuming a small emitting area. 
Only for IRAS 18151 and IRAS 18089, the map intensities are substantially lower ($\approx$40\% of the pointed observations) for unknown reasons.
\new{In such cases, the pointed observations are more reliable, since their calibration is more thorough, with multiple references and longer integrations.}
We conclude that mapping modes are useful to measure source sizes, but usually underestimate line intensities, sometimes substantially.

\begin{table*}[t]
\caption{Observed and deconvolved source sizes (arcsec).} 
\label{t:sizes}
\begin{tabular}{lcccccc}
  \hline \hline
\noalign{\smallskip}
Source$^a$        & Major axis & Minor axis & Position angle (degrees) & Deconvolved$^b$ & Model$^c$ & 850--870\,\mic$^d$ \\
\noalign{\smallskip}
\hline
\noalign{\smallskip}

     IRAS 05358 &  12.0 (0.5) &  9.3 (0.4) &   -46 ( 6) &  4.4 & 15.0 & 30.0 \\
     IRAS 16272 &  15.9 (0.7) & 11.8 (0.6) &   -34 ( 5) &  8.8 &  18.0 & 50.0 \\
  NGC 6334 I(N) &  17.2 (0.5) & 12.5 (0.4) &    38 ( 3) & 11.3 & 21.6 & 42.0 \\
          W43 MM1 &  15.8 (1.6) & 12.8 (1.2) &   -89 (15) & 15.5 & 11.8 & 27.0 \\
          DR21(OH) &  12.4 (0.4) & 11.1 (0.4) &    -8 (11) &  5.4 & 28.8 & 33.0 \\
            W3 IRS5 &  14.1 (0.3) & 12.0 (0.3) &    -9 ( 5) &  7.8 & 40.0 & 57.0 \\
     IRAS 18089  &  11.0 (0.9) &  9.5 (0.8) &   -37 (21) & -0- &  ... & 17.0 \\
                W33A &  11.6 (0.7) & 10.0 (0.6) &   -44 (16) &  5.1 &  ... & 30.0 \\
         AFGL 2591 &  11.3 (0.6) &  9.7 (0.5) &    76 (12) &  3.8 & 25.2 & 25.2 \\
           G327-0.6 &  14.3 (0.5) & 10.2 (0.4) &    67 ( 4) &  5.5 &  ... & 24.0 \\
        NGC 6334I &  16.1 (0.5) & 15.0 (0.4) &    88 (13) & 12.6 &  ... & 40.0 \\
              G29.96 &  10.7 (0.3) &  9.6 (0.3) &   -13 (11) &  3.0 &  ... & 16.0 \\
              G31.41 &   9.9 (0.9) &  7.5 (0.7) &    -5 (13) & -0- &  ... & 15.0 \\
                G5.89 &  10.9 (0.2) & 10.0 (0.2) &   -29 ( 7) &  3.9 &  ... & 28.0 \\
              G10.47 &  11.9 (0.7) & 10.2 (0.6) &   -47 (14) &  5.5 &  ... & 10.0 \\
              G34.26 &  22.4 (1.0) & 16.9 (0.8) &   -17 ( 4) & 16.9 &  ... & 25.0 \\
           W51N-e1 &  14.5 (0.7) & 10.9 (0.5) &   -63 ( 6) &  8.2 &  ... & 27.0 \\

\noalign{\smallskip}
\hline
\end{tabular}
\tablefoot{$^a$: The source fitting failed for IRAS 18151 and NGC 7538.
$^b$: Equivalent circular axis; a value of zero means that the source is unresolved.
$^c$: From \citet{chavarria2010} for W3 IRS5, Herpin (priv. comm.) for AFGL 2591; other sources from \citet{herpin2016}.
$^d$: 3$\sigma$ radii from \citet{vandertak2013}.
}
\end{table*}

\section{Discussion}
\label{s:disc}

\subsection{Origin of \hho\ and \hhoe\ line emission and absorption}
\label{ss:hhoe}

\begin{figure*}[tb]
\centering
\includegraphics[width=9cm,angle=0]{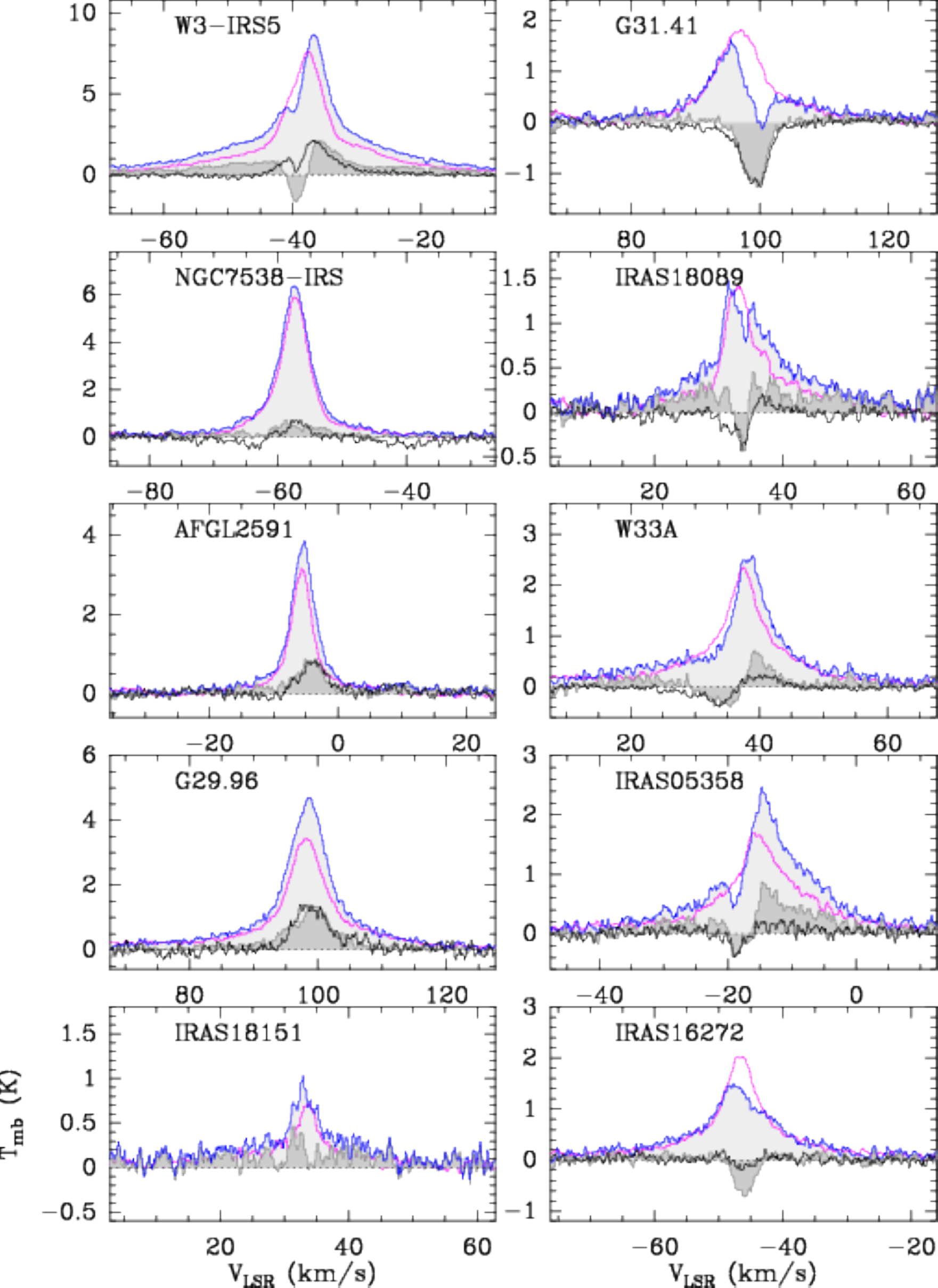}
\includegraphics[width=9cm,angle=0]{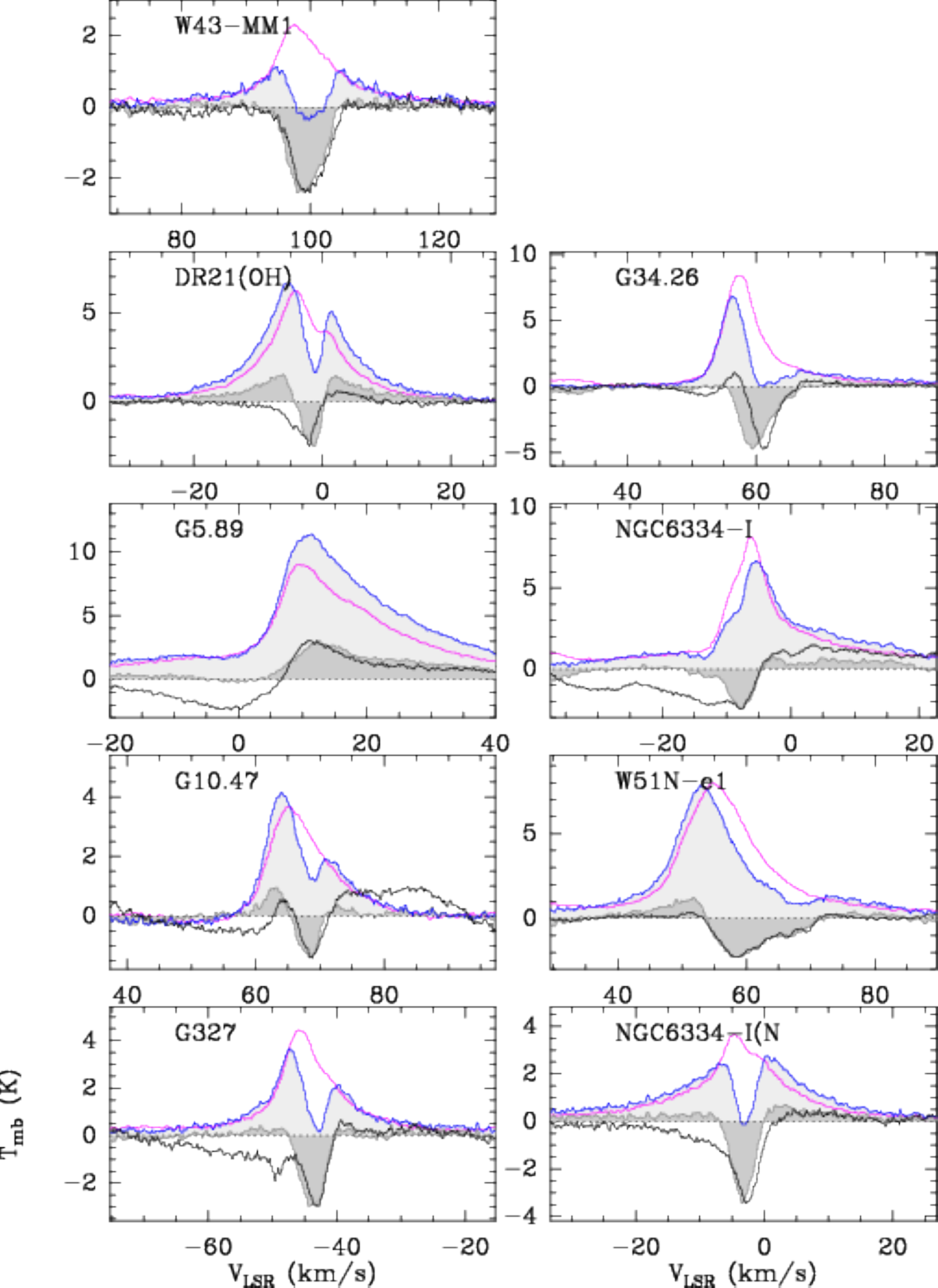}
\caption{Spectra of the \hho\ 987 and 752 GHz lines (blue and purple histograms), and their difference (shaded grey histogram), compared with the \hhoe\ line profile (black) histogram.}
\label{f:noscaling}
\end{figure*}

Figure~\ref{f:noscaling} compares the observed \hhoe\ line profiles with those of the \hho\ 987 and 752 GHz lines. 
For the 987 GHz line, we use the pointed observations \newer{rather than convolving the map data}, because of the calibration issue with the maps (\S\ref{ss:maps}) and because the map data have higher noise levels.
Remarkably, the \hhoe\ line profile (shown in \new{black}) is very similar to the {\it difference} between the two \hho\ lines (shown in grey).
As found before for the case of W43-MM1 \new{by another method} \citep{jacq2016},
this close similarity implies that the \hhoe\ absorption originates in warm gas ($T \gtsim 100$\,K).
Given the upper level energies of the two \hho\ lines (101 and 137\,K), the bulk of the \hhoe\ absorption must arise in gas with temperatures between $\sim$100 and $\sim$140\,K.
These temperatures are just above the point where \hho\ ice sublimates from dust grains, which is expected to lead to a strong increase in the gas-phase \hho\ abundance \citep{boogert2015}.
The \hhoe\ absorption is unlikely to arise in the cold outer envelope, where the \hho\ abundance is too low to create detectable absorption in \hhoe\ (cf.\ \citealt{shipman2014}).
\newer{The success of the subtraction procedure shows that the outer envelope does not contribute to the \hhoe\ absorption.}

For the sources W3 IRS5, NGC 7538, W33A, AFGL 2591, G29.96, G10.47, and W51N, the subtraction also reproduces \hhoe\ \textit{emission} features.
Since emission is sensitive to beam filling factors, this similarity is even stronger evidence that the \hhoe\ line originates between the layers where the 752\,GHz line is excited and where the 987\,GHz line is excited.
In the models by \citet{vandertak2013}, this zone occurs typically at radii of 1000-5000 AU, depending on the luminosity of the source. 
This region is small enough that it is often difficult to observe (\newer{e.g., 2--10$''$ diameter} at a distance of 1\,kpc).

\begin{figure}[tb]
\centering
\includegraphics[width=9cm,angle=0]{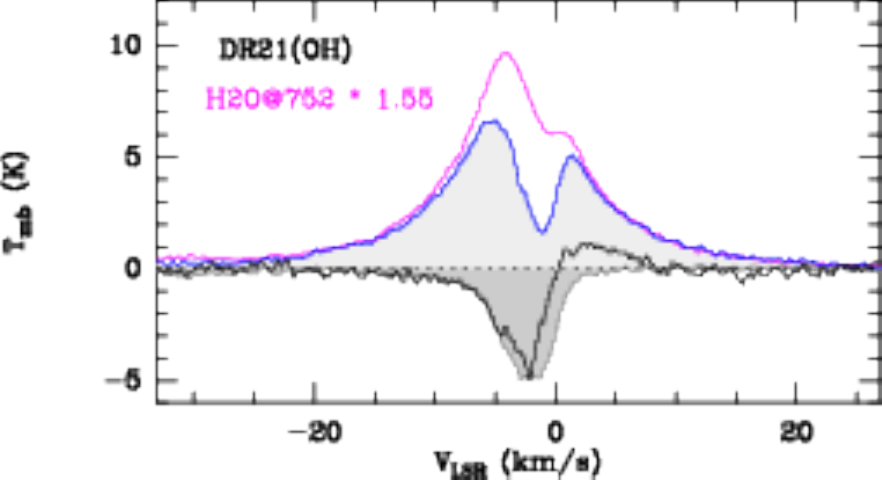}
\caption{As previous figure, for DR21(OH), with the 752 GHz profile scaled to optimize the match to the \hhoe\ line wings. 
}
\label{f:scaling}
\end{figure}

In some cases, scaling the 752\,GHz profile before subtracting it from the 987\,GHz line profile improves the match of the difference to the \hhoe\ profile (Fig.~\ref{f:scaling}), in particular for the line wings.
The scaling factors that best match the observed profiles range from $\approx$1 for sources with small deconvolved sizes (Table~\ref{t:sizes}) to $\approx$1.8 for the most extended sources.
These values are just as expected from beam size differences between the 987 and 752 GHz spectra, assuming equal excitation temperatures. 
There may be other pairs of lines whose differences enable us to probe specific layers of the protostellar cores.  

Toward several of the more massive sources, the \hhoe\ line profiles show absorption in the line wings, especially on the blue-shifted side. 
Clearly, the \hhoe\ column density is sufficient to absorb even at velocities only seen in the wings.
The \ceo\ 9--8 spectra show no such high-velocity signals, which implies that the \hho\ abundance is enhanced in the high-velocity gas \citep{herpin2016}.
For example, the \hhoe\ spectrum toward NGC 6334 I(N) shows absorption out to at least 15--20 \kms\ from line center, which has no counterpart in \ceo.
For this source, the integrated \hhoe\ absorption from the envelope (roughly between --6 and +1 \kms, \newer{which has a counterpart in \ceo\ emission}) is approximately equal to that in the high-velocity blue wing.
In contrast, the \ceo\ 9--8 line indicates $\gtsim$10$\times$ less mass at high velocities, implying an \hho\ abundance enhancement by more than an order of magnitude. 

Similar conclusions hold for the other sources, except for G5.89 and G34.26 where weak wings are seen on the \ceo\ 9--8 profiles.
The lack of high-velocity \ceo\ 9--8 emission for most sources is not an excitation effect, as low-$J$ \ceo\ lines do not show wings either \citep{hatchell1998,watson2003,gibb2004,thomas2007}.
We conclude that \hho\ abundances in high-mass protostellar outflows are $\gtsim$10$\times$ higher than in the envelopes.

\new{The \hho\ abundance in these sources thus appears to have at least 3 levels: low in the outer envelope, high in the inner envelope, and very high in the outflow.}
\newer{This is in line with the work of \citet{vandertak2010}, who used HIFI maps of the DR21 region in \thco\ 10--9 and \hho\ $1_{11}$--$0_{00}$ to derive \hho\ abundances of $\sim$10$^{-10}$ for the cool outer envelope, $\sim$10$^{-8}$ for the warm inner envelope, and $\sim$10$^{-6}$ for the shocked outflowing gas.}

\subsection{Infall rates and trends}
\label{ss:mdot}

The rightmost column of Table~\ref{t:pars} gives estimates of the infall rates onto our sources. These were calculated using 
$$ \dot{M}_{\rm acc} = 4 \pi R^2 \: m({\rm H}_2) \: n({\rm H}_2) \: | V_{\rm inf} | $$
where $m({\rm H}_2)$ is the mass of the \hh\ molecule, 
and the \new{absolute value of the} infall speed $V_{\rm inf}$ is taken from Table~\ref{t:pars}.
\new{Infall motion appears negative as gas is moving toward the center of the reference frame of our models.} 
Given the similarity of the \hhoe\ absorption profile with the difference of the \hho\ 987 and 752 GHz profiles (\S~\ref{ss:hhoe}), we adopt the radius of the 120~K point in the envelope models from \citet{vandertak2013} for $R$, and the density at that radius for $n({\rm H}_2)$. 
These radii vary between 800 and 9000~au, and the densities from \pow{7}{5} to \pow{5}{7}~\ccm.
Our observed (deconvolved) sizes agree well (within a factor of 2) with the upper end of this range, except for W43~MM1, G10.47, and W51N, \newer{where the observed values are larger.}

The resulting infall rates (Table~\ref{t:pars}, right column) are seen to range from $\sim$\pow{7}{-5} to $\sim$\pow{2}{-2} \msol/yr.
These values are in reasonable agreement with other observations \citep[e.g.,][]{koenig2017} and with theoretical models \citep{tan2014,motte2017}.
They should be considered order of magnitude estimates, because of our simplified treatment assuming spherical symmetry.
\new{The observational uncertainty through the measured line velocities is only a $\sim$30\% effect.}
The derived infall rates depend only weakly on the adopted radius and density: the envelopes of our sources have density profiles which drop off approximately as $R^{-2}$, so that the effects of $R$ and $n$ on $\dot{M}$ tend to cancel each other.

\begin{figure}[tb]
\centering
\includegraphics[width=9cm,angle=0]{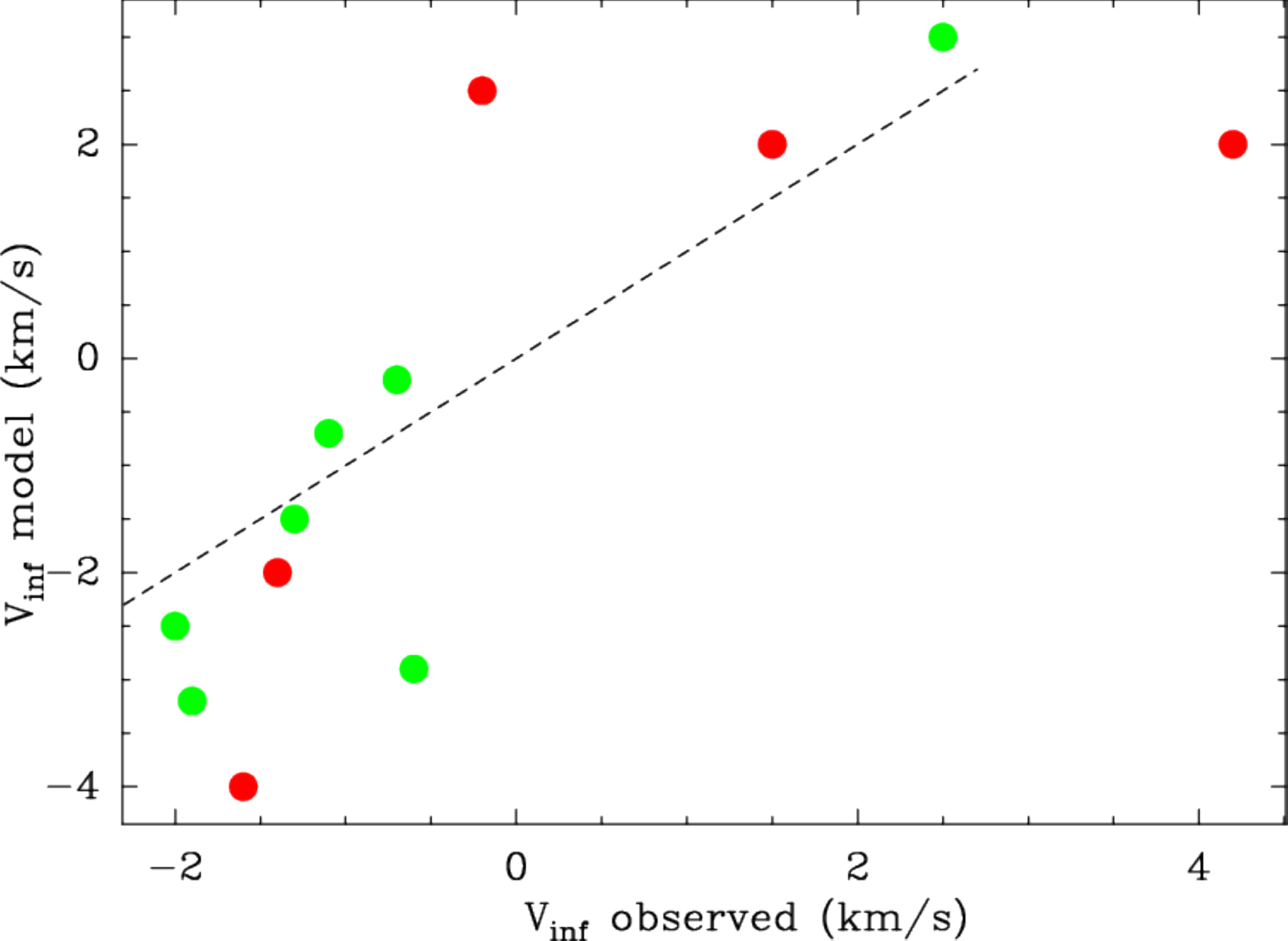}
\caption{\new{Infall} velocities estimated from peak shift between \hhoe\ and \ceo\ lines versus values from detailed fits to \hho\ line profiles (using RATRAN) by \citealt{herpin2016} \new{(green)} and \citealt{choi2015} \new{(red)}. \new{The dashed line denotes 1:1 correspondence}.}
\label{f:models}
\end{figure}

For our subsamples of mid-infrared quiet \newer{and --bright HMPOs}, \citet{herpin2016} \newer{and \citet{choi2015}} have made detailed models of the \hho\ distribution in the protostellar envelope, including simple step functions for the \hho\ abundance in the inner and outer envelope.
In order to fit the line profiles of \hho, \hhoe, and \hhos\ in the pointed HIFI spectra, Herpin \newer{and Choi} had to include radial motions in their models.
Figure~\ref{f:models} compares their derived \new{infall} velocities to the values found here; the \newer{dashed line indicates 1:1 correspondence}.
The two types of estimates of the radial velocity are seen to agree qualitatively, both in the sign of the velocity (infall or outflow) and in its magnitude.
The simple estimates of the inflow / expansion velocity are on average $\approx$2$\times$ lower than those from the detailed models, although for four sources, they are actually larger.
We consider this agreement as \newer{reasonable, given the differences between the two approaches.}

Since our source sample covers a range of luminosities, envelope masses, and evolutionary stages, we have investigated if our derived source sizes, infall velocities, and infall rates show any trends with \lbol, \menv, and age. 
To estimate the \newer{relative} ages of our sources, we have used the ratio of \lbol/\menv, which is straightforward to compute and appears to be a robust age tracer \citep{molinari2016}.
In addition, we use the presence of hot molecular cores and/or ultracompact HII regions as a sign of a relatively evolved stage.
Third, we have looked for trends with the virial mass and the ratio \mvir/\menv, \newer{proposed} as a stability parameter by \citet{koenig2017}.
\newer{Virial masses are calculated following \citet{giannetti2014}, using line widths from Table A.2 of \citet{vandertak2013}.}

\new{The only significant trend that we find is between the linear sizes of our sources with their virial masses. 
Since virial mass depends on size, this trend probably just means that the line width is similar for all sources. 
In addition, the infall rates seem to increase with virial mass and with the evolutionary indicator \lbol/\menv, but the statistical significance of these trends is small. 
Even the relation with \mvir\ has a Pearson correlation coefficient of only $r=$0.58.
For a sample size of $N$=11, this $r$-value corresponds to a probability of false correlation of $p=$6\%, i.e. a $\approx$2$\sigma$ significance. 
We conclude that the accretion rates may increase with circumstellar mass and with evolutionary stage, but that larger source samples are required to test these claims.
}

\section{Conclusions}
\label{s:concl}

Based on our measured velocity shifts between \hhoe\ absorption and \ceo\ emission, infall motions appear to be common in the embedded phase of high-mass star formation, with typical accretion rates of $\sim$\pow{1}{-4} \msol/yr.
We find a tentative trend that the highest accretion rates occur for the most massive sources, which is globally consistent with current models of high-mass star formation \citep{tan2014,motte2017}.
Our data do not allow to distinguish between such models, though.

In addition, the accretion rates may increase with age, unlike in the low-mass case, where accretion rates drop from the Class 0 to the Class III stage, and \newer{are} highly episodic \citep{dunham2014}. 
Signs of episodic accretion, which is well established in the low-mass case, have recently been reported for a high-mass star, in the form of mid-infrared variability suggesting accretion `bursts' \citep{caratti2017}.

Our data do not allow to discern trends within specific types of sources, nor with protostellar luminosity.
A study of \hho\ line profiles toward a large ($N\sim 100$) sample is needed to distinguish such trends and to search for episodic behaviour.
\newer{Data from the Herschel open time programs by Bontemps and Wyrowski may be suitable for this purpose. 
In the future,} such studies will be possible with ESA's SPace Infrared telescope for Cosmology and Astrophysics (SPICA)\footnote{\tt http://www.spica-mission.org/} \citep{roelfsema2018,vandertak2018} around 2030, and NASA's Origins Space Telescope (OST)\footnote{\tt https://asd.gsfc.nasa.gov/firs/} \new{\citep{battersby2018}} around 2040.

\begin{acknowledgements}

This paper is dedicated to the memory of Malcolm Walmsley, who passed away on 1 May 2017 at the age of 75. 
We remember Malcolm as a great source of inspiration, and we will miss his sharp insight and kind manner. 
\new{The authors thank the WISH team led by Ewine van Dishoeck for inspiring discussions, and the anonymous referee for useful comments on the manuscript.}
This research has used the following databases: ADS, CDMS, JPL, and LAMDA.


\par HIFI was designed and built by a consortium of institutes and university departments from across Europe, Canada and the US under the leadership of SRON Netherlands Institute for Space Research, Groningen, The Netherlands with major contributions from Germany, France and the US. Consortium members are: Canada: CSA, U.Waterloo; France: CESR, LAB, LERMA, IRAM; Germany: KOSMA, MPIfR, MPS; Ireland, NUI Maynooth; Italy: ASI, IFSI-INAF, Arcetri-INAF; Netherlands: SRON, TUD; Poland: CAMK, CBK; Spain: Observatorio Astron\'omico Nacional (IGN), Centro de Astrobiolog\'{\i}a (CSIC-INTA); Sweden: Chalmers University of Technology - MC2, RSS \& GARD, Onsala Space Observatory, Swedish National Space Board, Stockholm University - Stockholm Observatory; Switzerland: ETH Z\"urich, FHNW; USA: Caltech, JPL, NHSC.
\end{acknowledgements}

\bibliographystyle{aa}
\bibliography{wish-hmpo}


\section*{Appendix: Maps of all sources}

\begin{figure*}
\includegraphics[width=18cm]{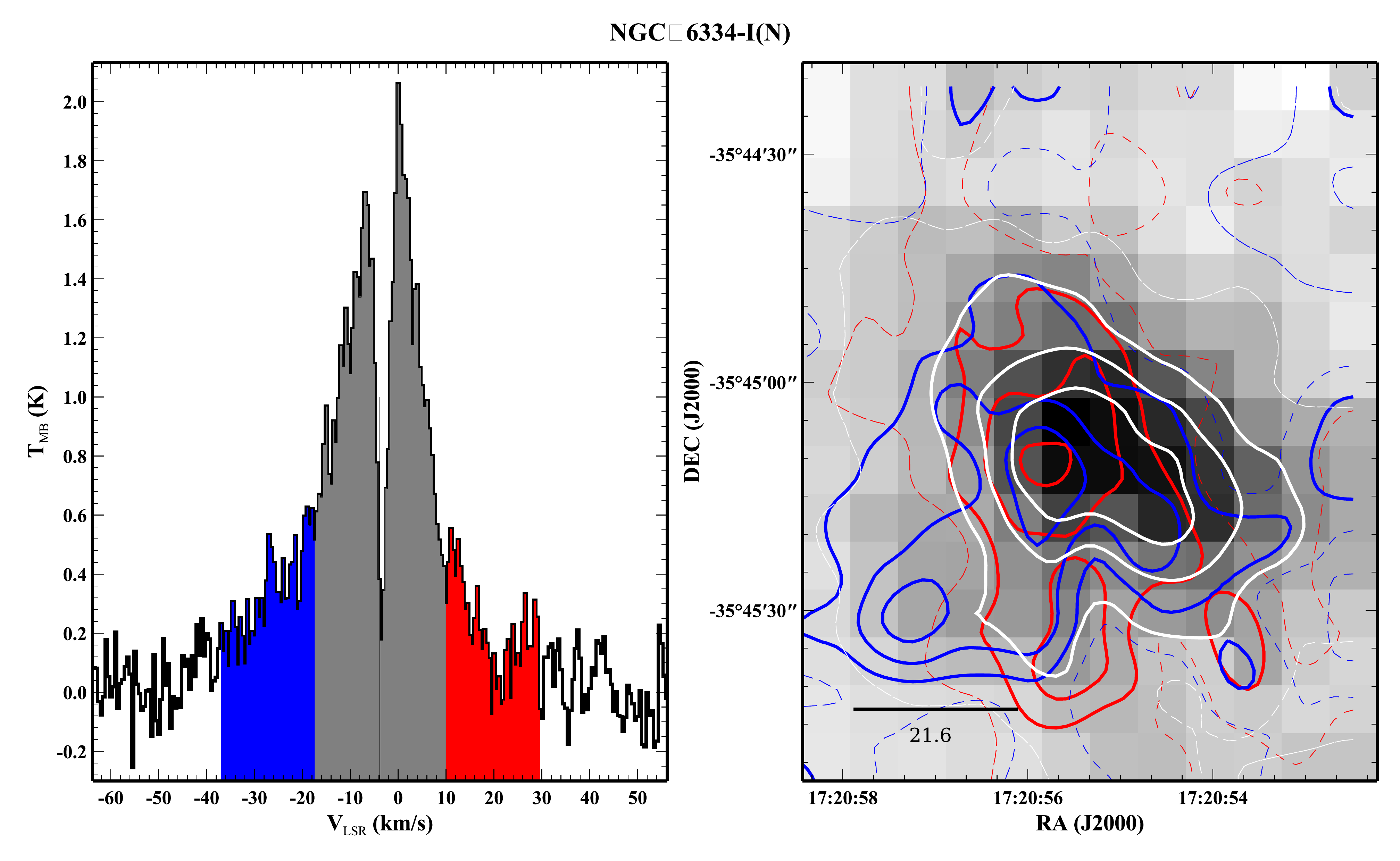}
\caption{As previous figure, for NGC 6334I(N)}
\end{figure*} 

\begin{figure*}
\includegraphics[width=18cm]{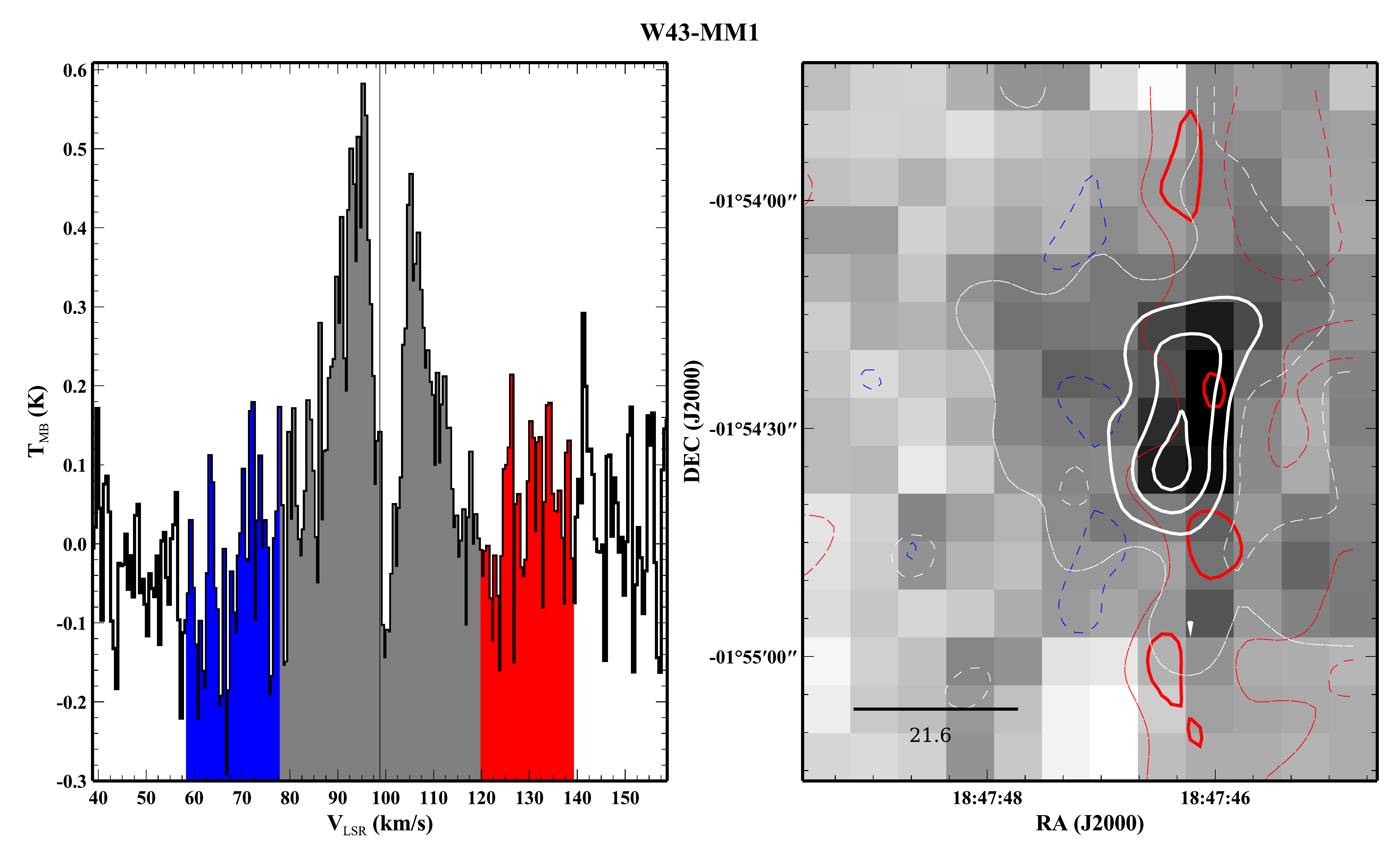}
\caption{As previous figure, for W43-MM1.}
\end{figure*} 

\begin{figure*}
\includegraphics[width=18cm]{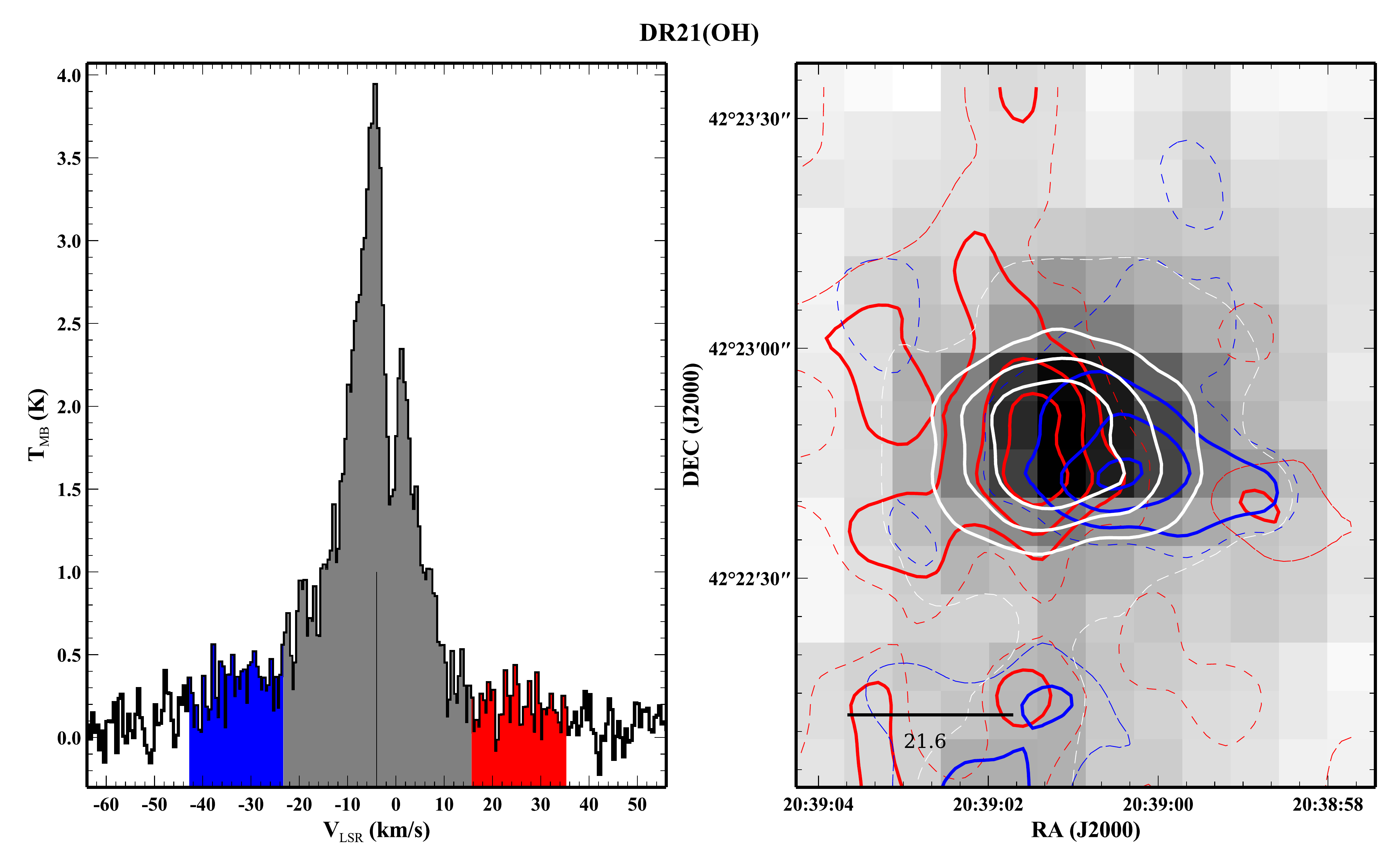}
\caption{As previous figure, for DR21(OH).}
\end{figure*} 

\begin{figure*}
\includegraphics[width=18cm]{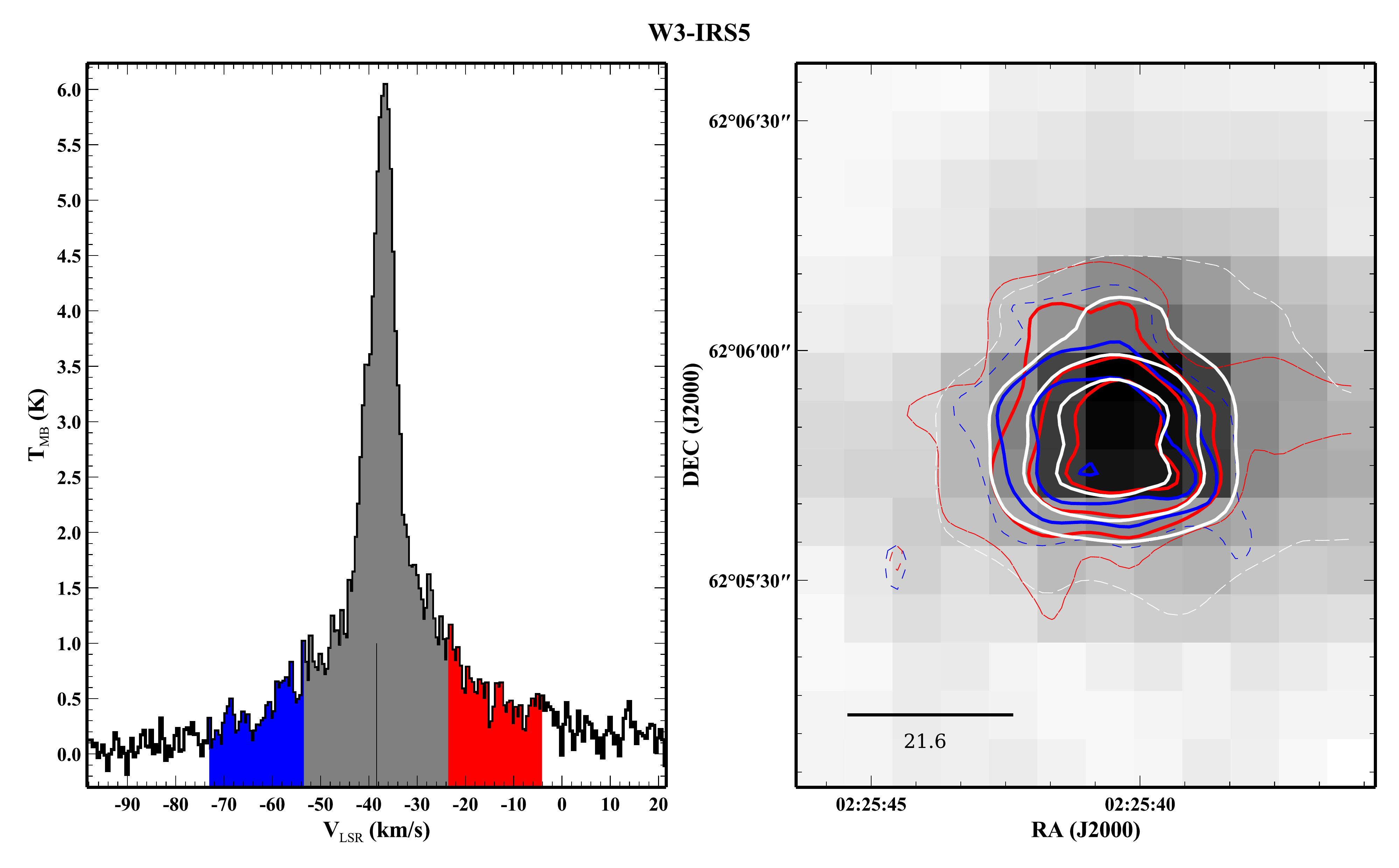}
\caption{As previous figure, for W3 IRS5.}
\end{figure*} 

\begin{figure*}
\includegraphics[width=18cm]{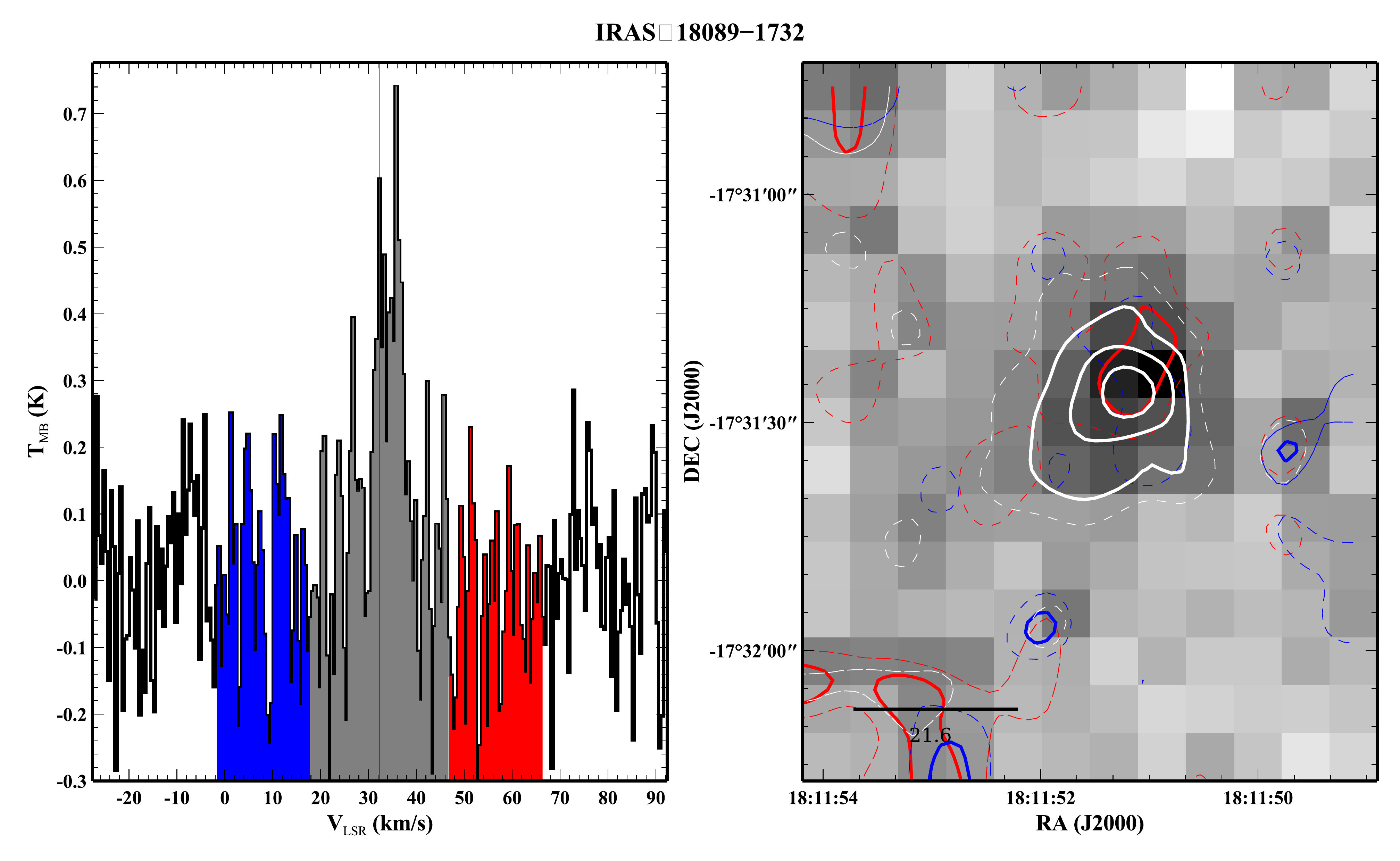}
\caption{As previous figure, for IRAS 18089.}
\end{figure*} 

\begin{figure*}
\includegraphics[width=18cm]{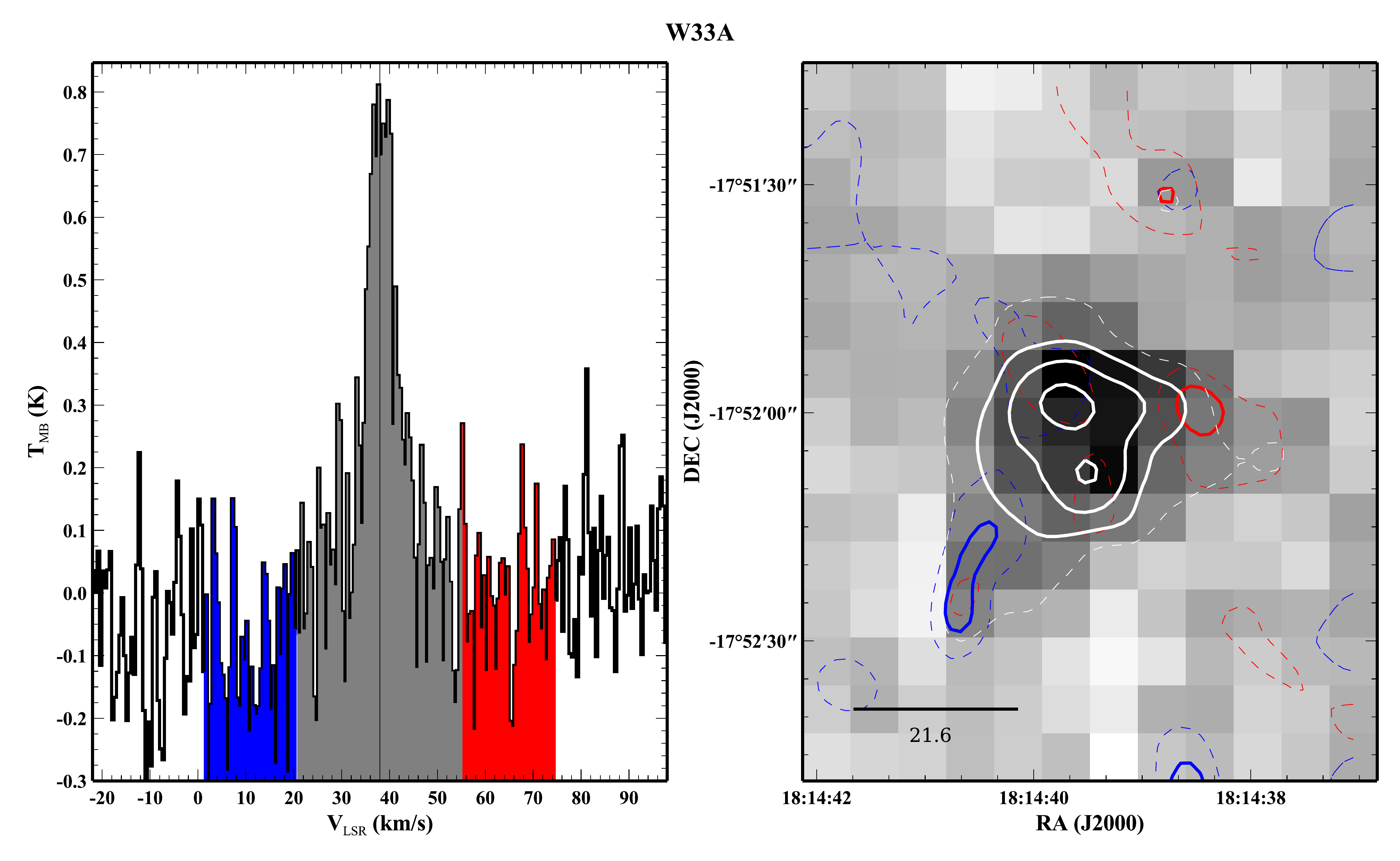}
\caption{As previous figure, for W33A.}
\end{figure*} 

\begin{figure*}
\includegraphics[width=18cm]{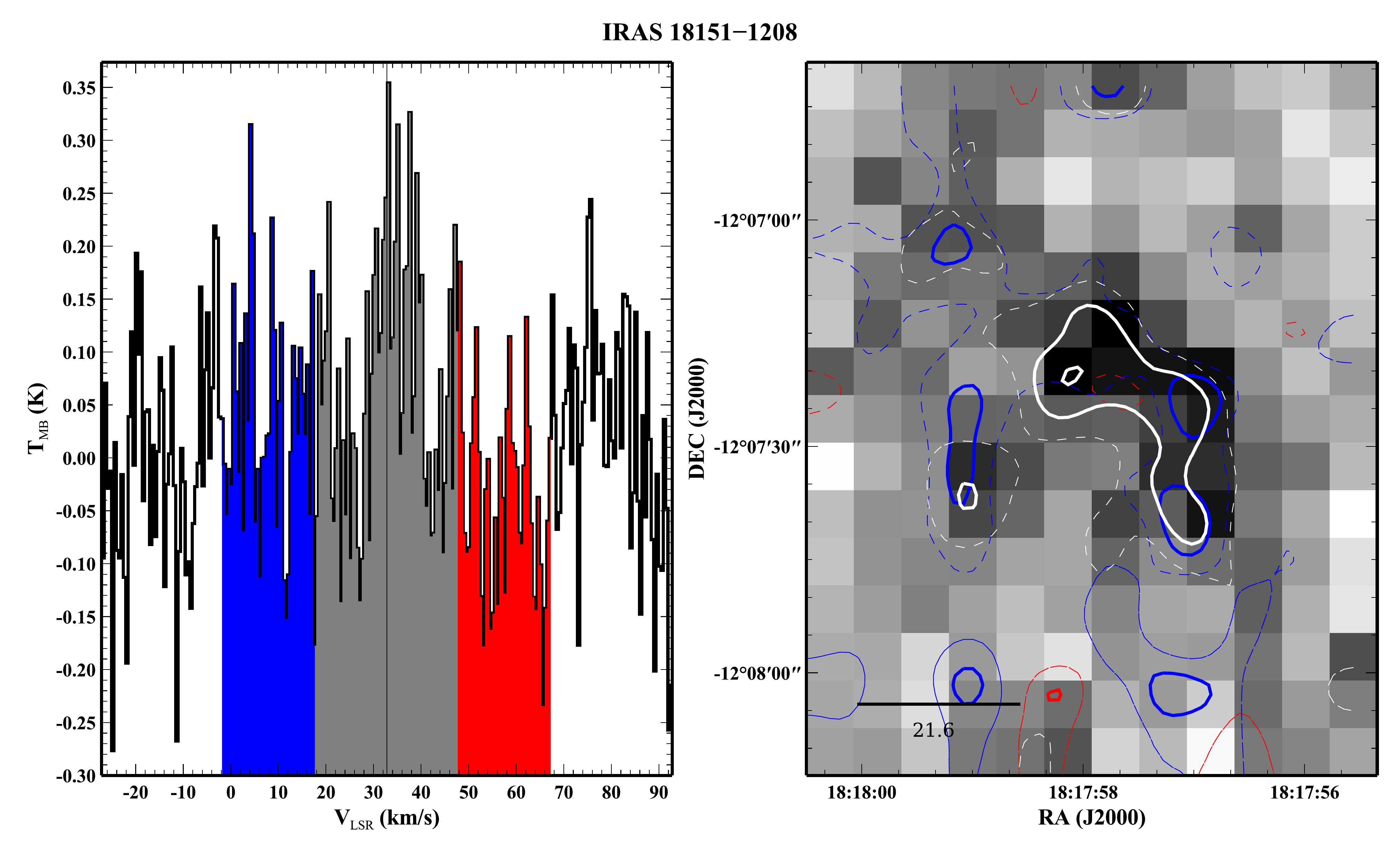}
\caption{As previous figure, for IRAS 18151.}
\end{figure*} 

\begin{figure*}
\includegraphics[width=18cm]{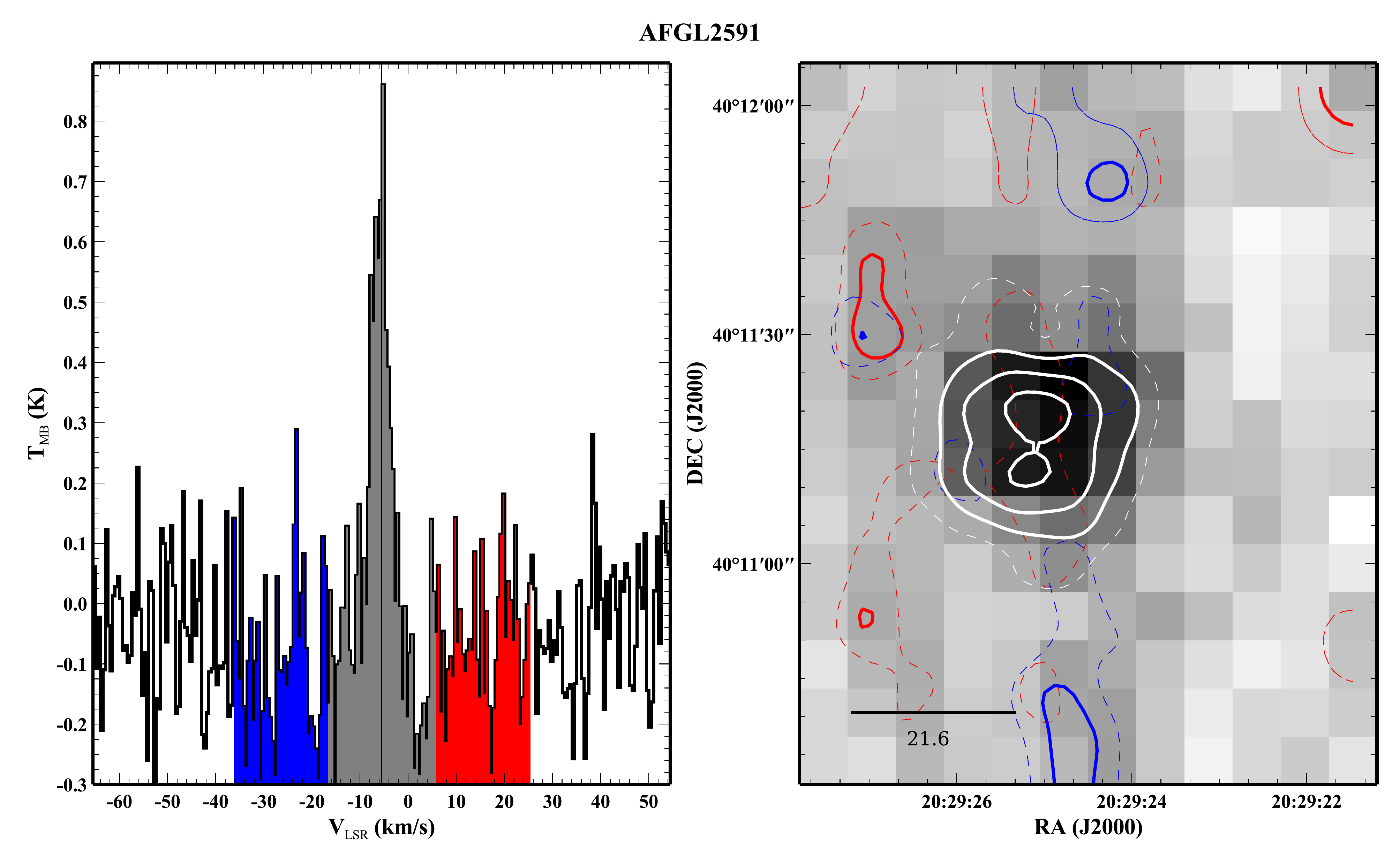}
\caption{As previous figure, for AFGL 2591.}
\end{figure*} 

\begin{figure*}
\includegraphics[width=18cm]{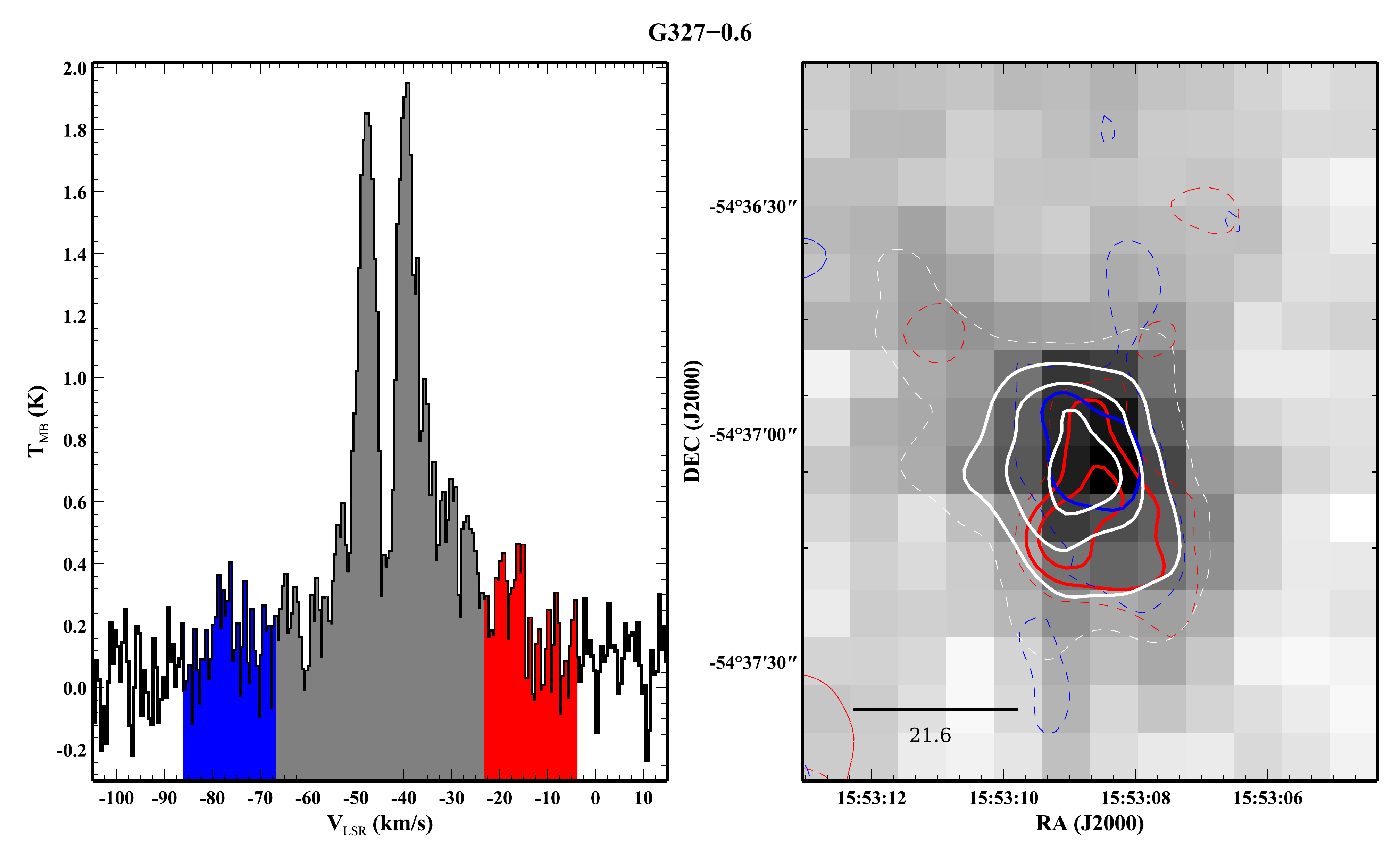}
\caption{As previous figure, for G327.}
\end{figure*} 

\begin{figure*}
\includegraphics[width=18cm]{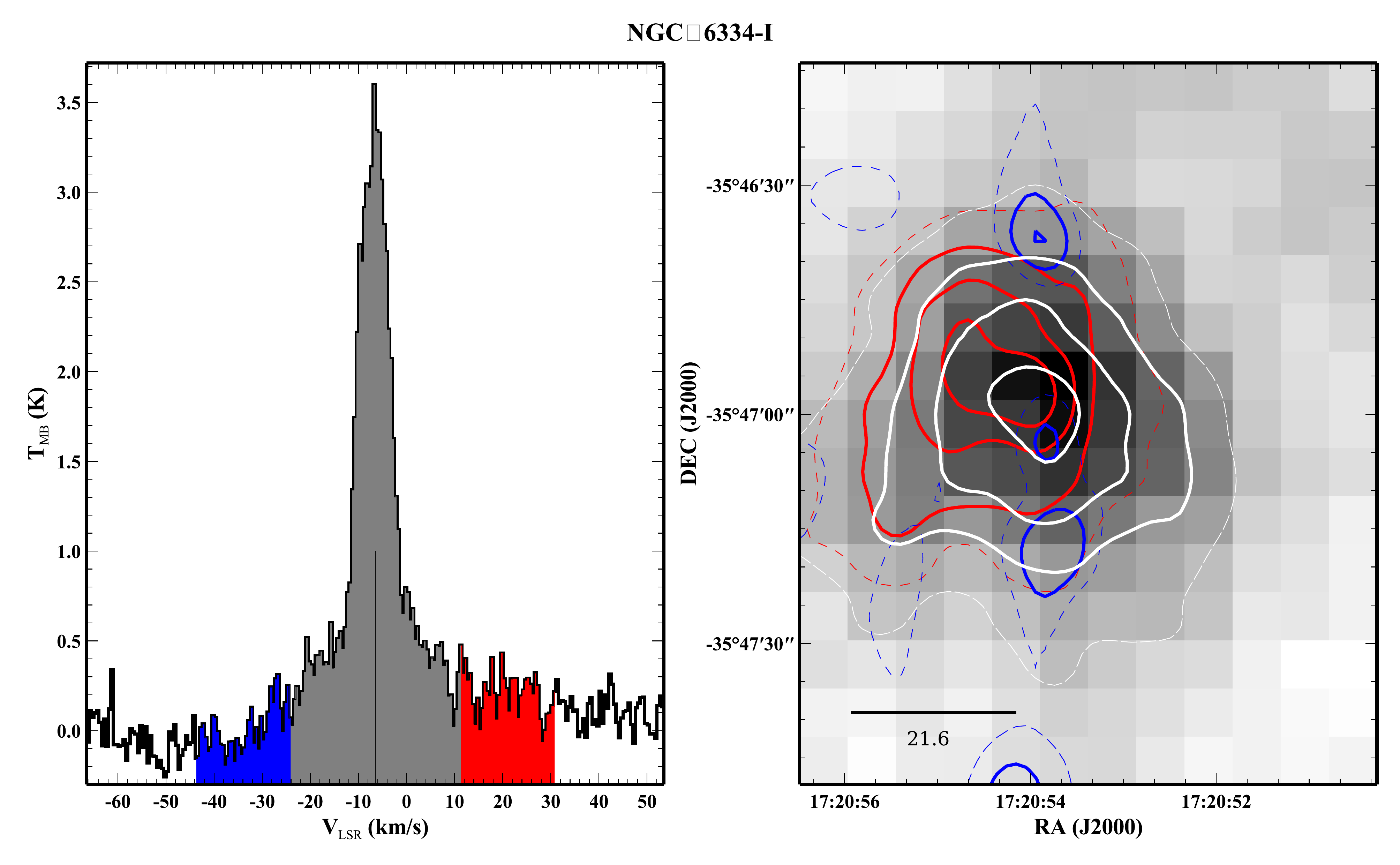}
\caption{As previous figure, for NGC 6334I.}
\end{figure*} 

\begin{figure*}
\includegraphics[width=18cm]{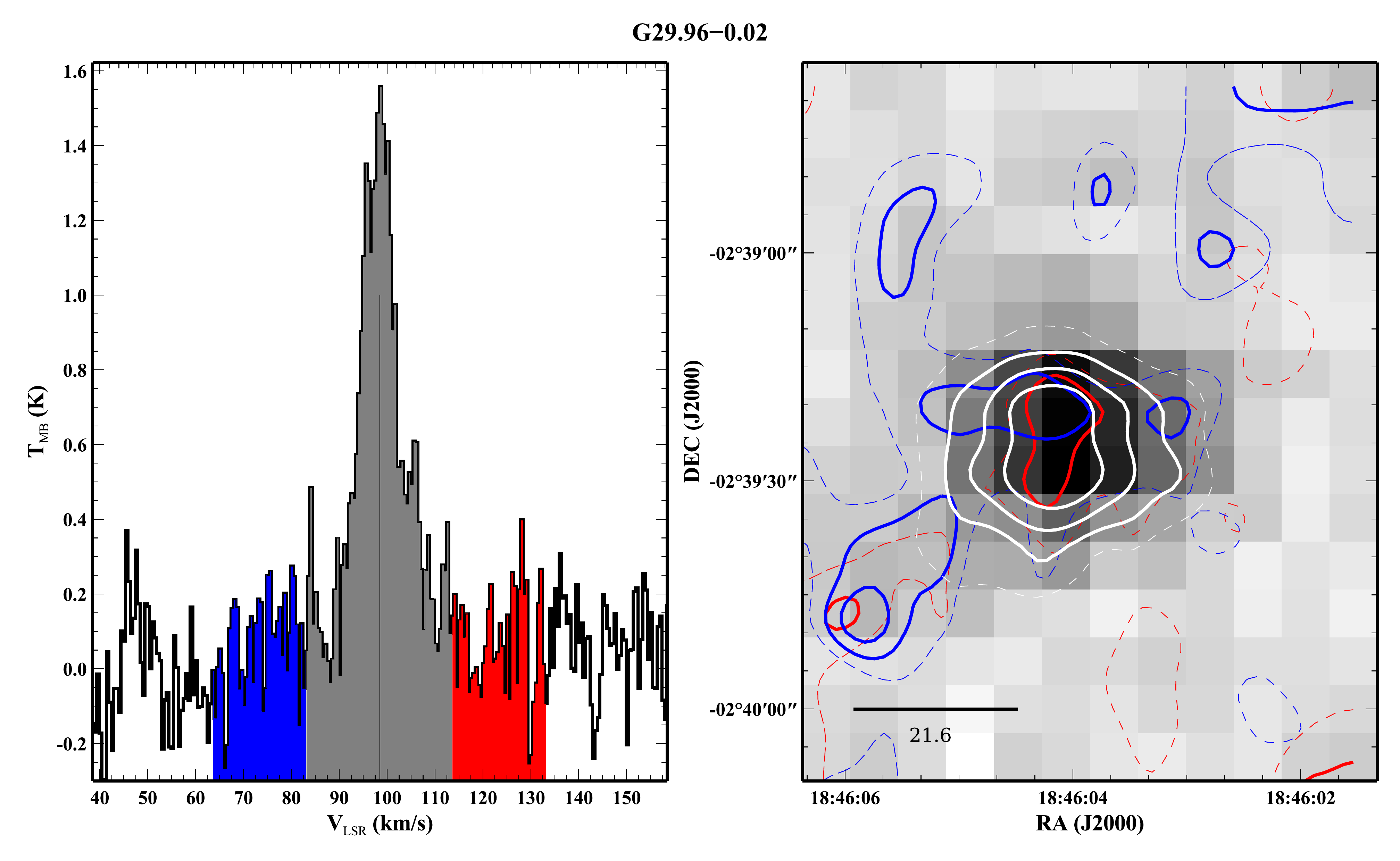}
\caption{As previous figure, for G29.96.}
\end{figure*} 

\begin{figure*}
\includegraphics[width=18cm]{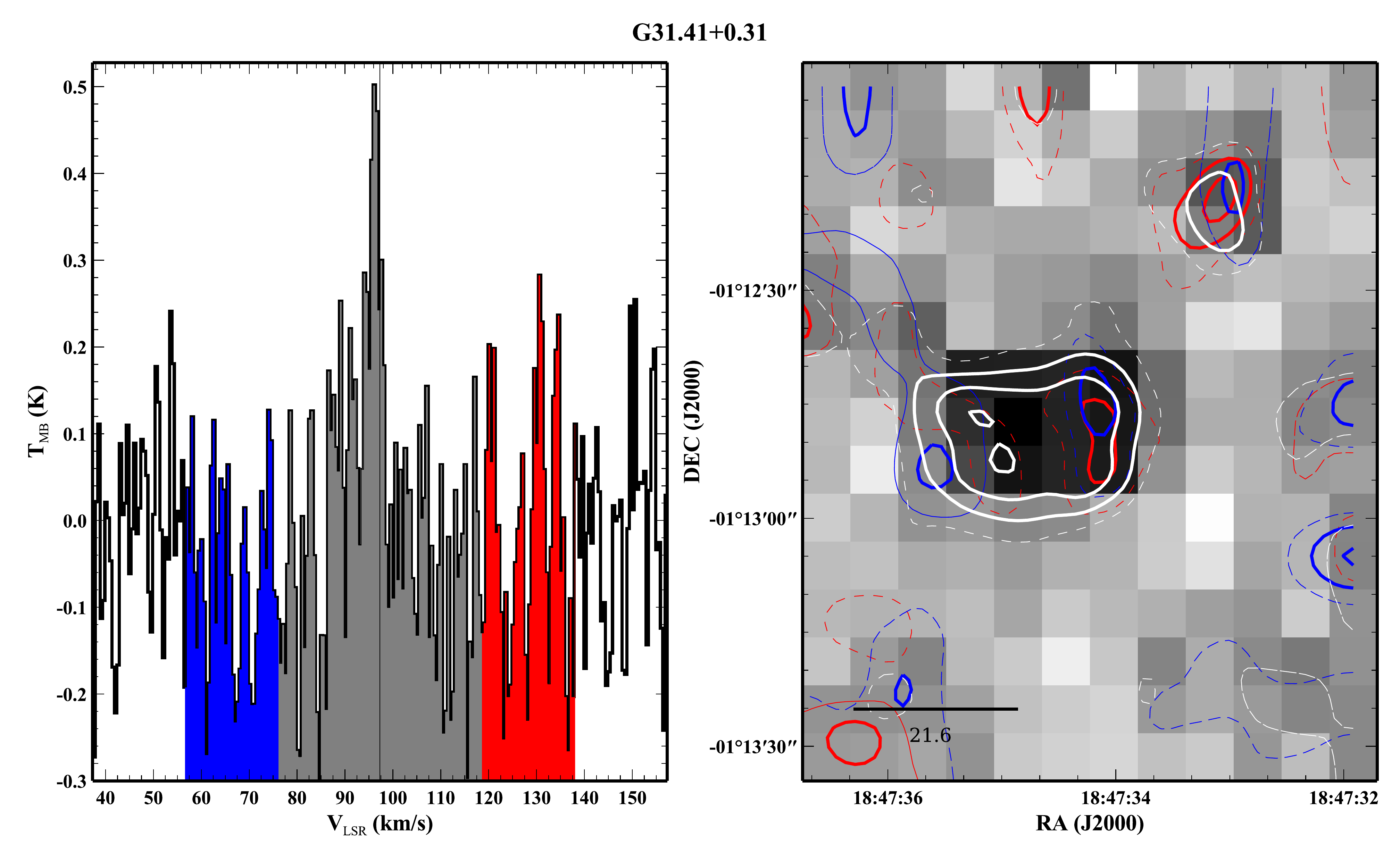}
\caption{As previous figure, for G31.41.}
\end{figure*} 

\begin{figure*}
\includegraphics[width=18cm]{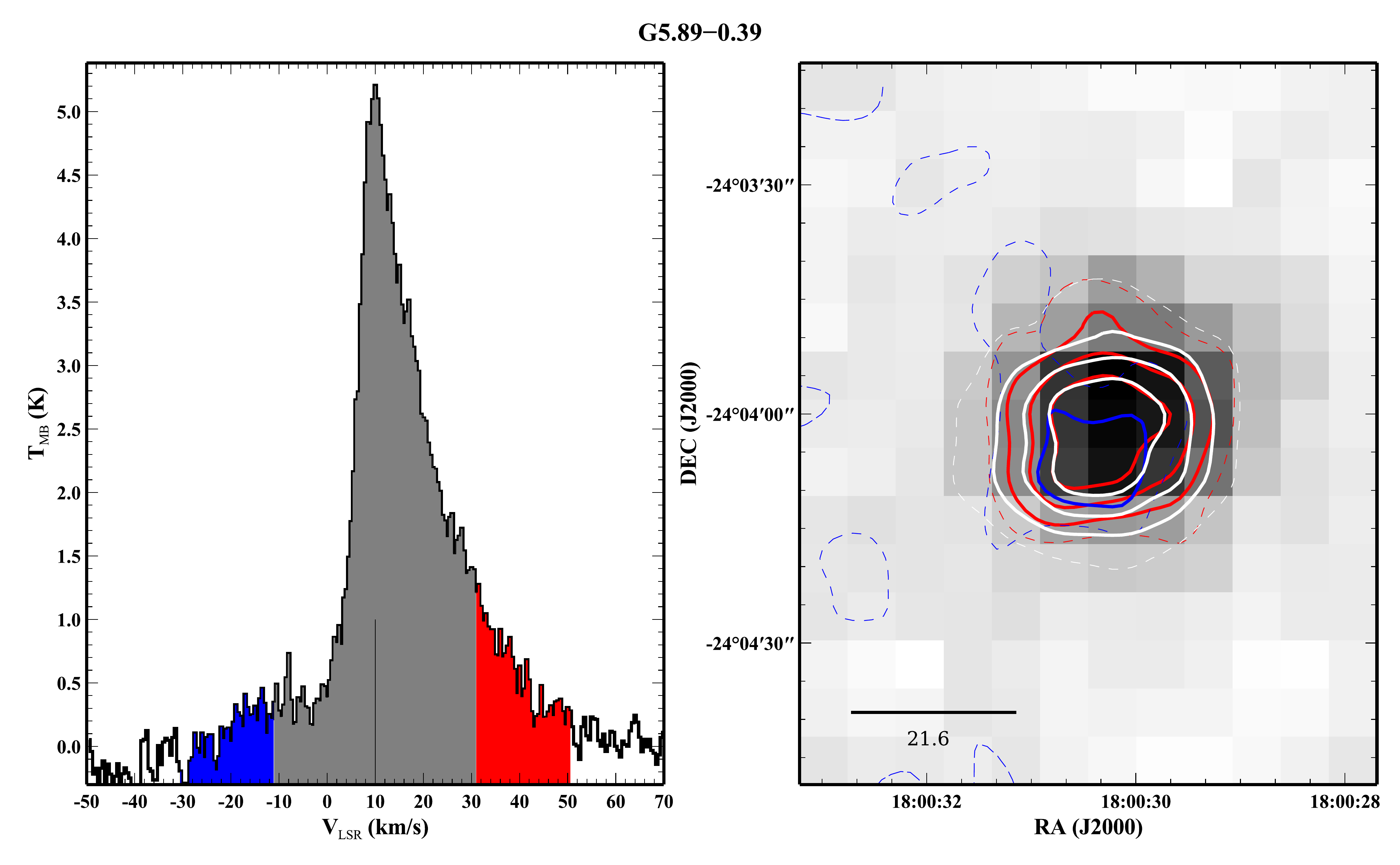}
\caption{As previous figure, for G5.89.}
\end{figure*} 

\begin{figure*}
\includegraphics[width=18cm]{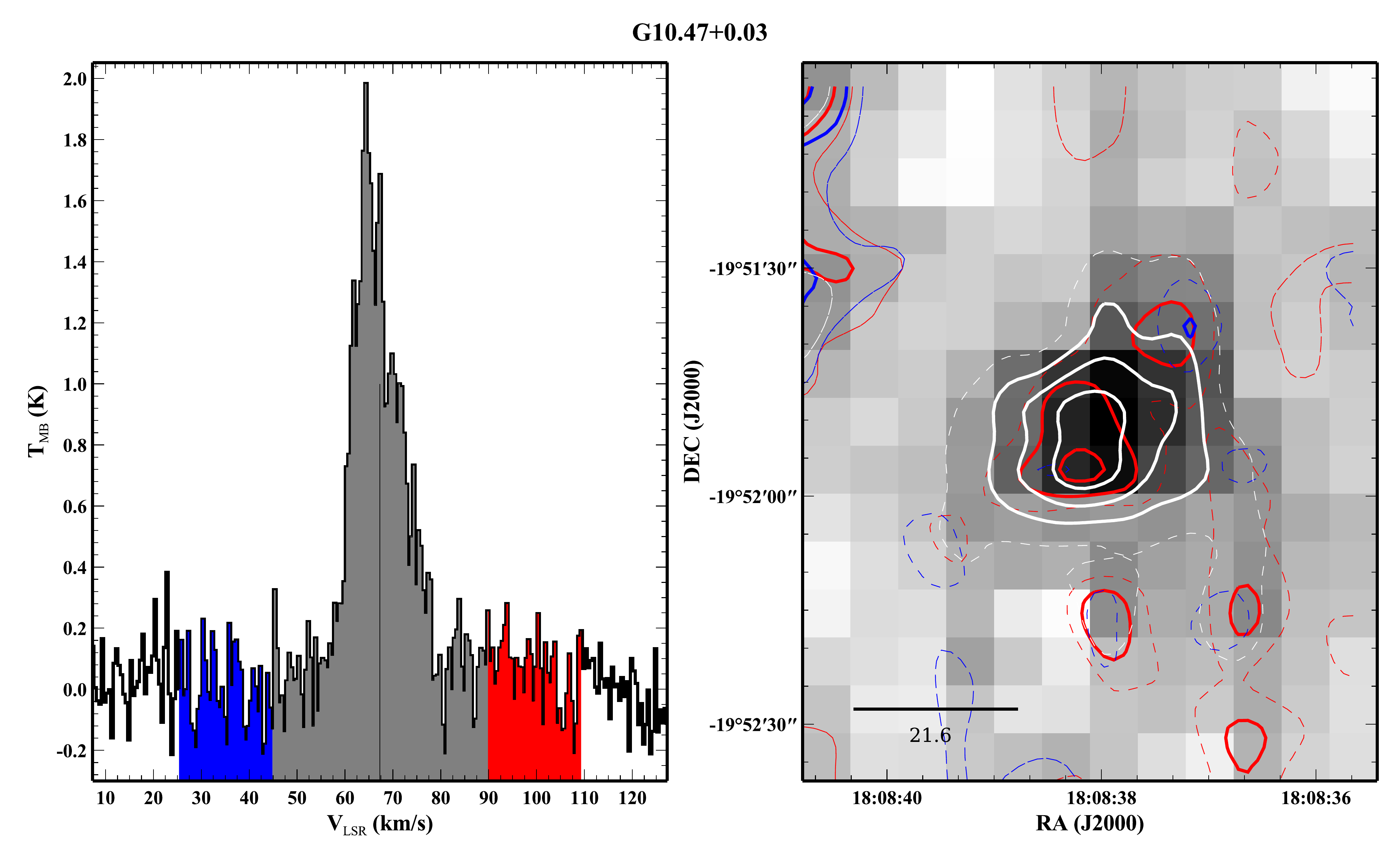}
\caption{As previous figure, for G10.47.}
\end{figure*} 

\begin{figure*}
\includegraphics[width=18cm]{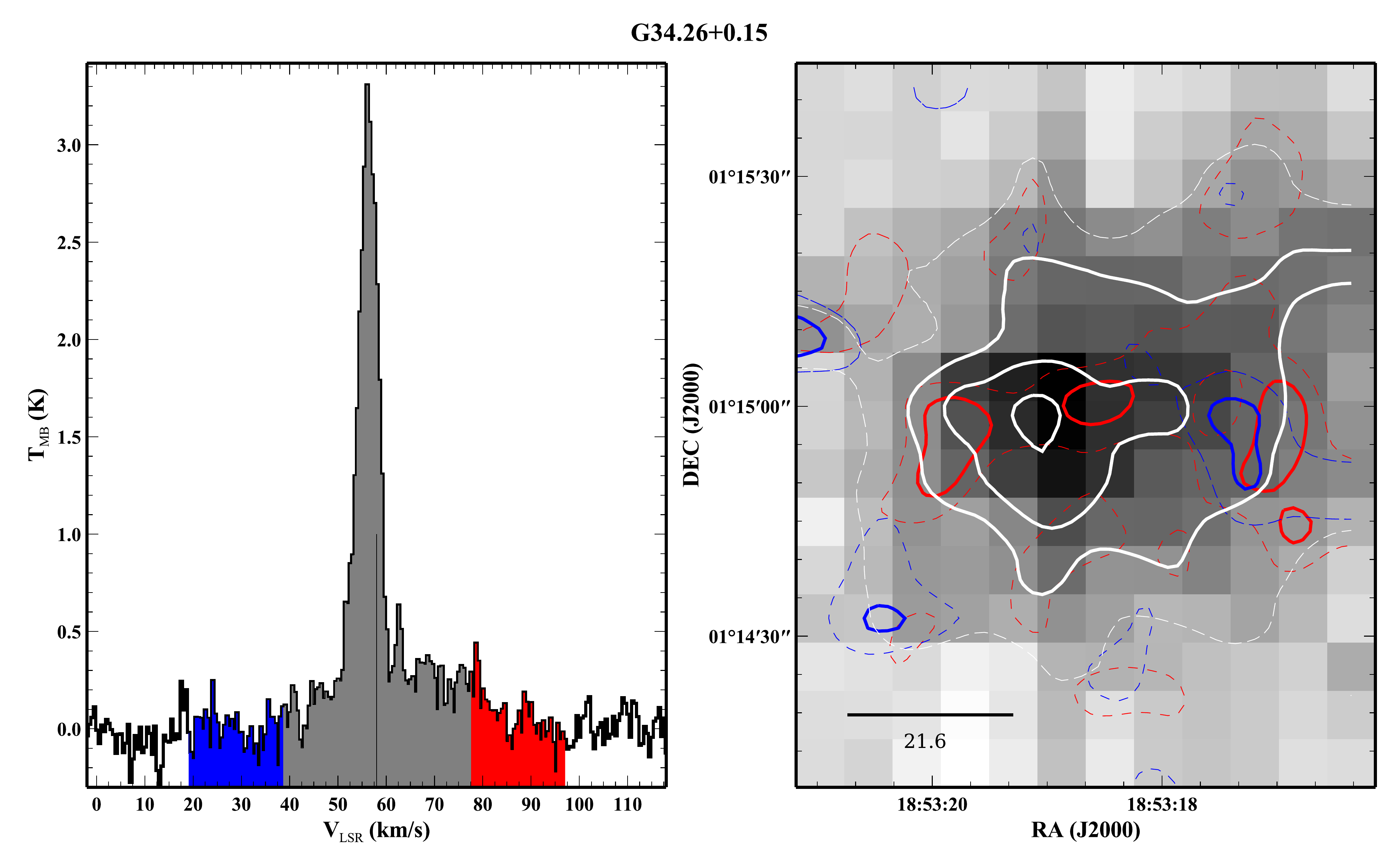}
\caption{As previous figure, for G34.26.}
\end{figure*} 

\end{document}